\documentclass[submission,copyright]{eptcs}

\providecommand{\event}{MARS 2020}

\usepackage{underscore}
\usepackage{breakurl}
\usepackage{graphicx}
\usepackage{subcaption}

\usepackage{xcolor}
\definecolor{grey}{cmyk}{0,0,0,0.8}
\hypersetup{bookmarksopen=false,colorlinks=true,linkcolor=grey,citecolor=grey,urlcolor=grey}

\usepackage[utf8]{inputenc} 

\usepackage[OT1]{fontenc}

\usepackage{fancyhdr} 
\usepackage{makeidx}
\usepackage{graphicx}
\usepackage{color}
\usepackage{array}

\ifx\hypersetup\undefined
\usepackage[hyperfootnotes=false]{hyperref}
\fi

\usepackage{breakurl} 
\usepackage{amssymb} 
\usepackage{amsfonts}

\ifx\keywords\undefined
\newcommand{\keywords}[1]{\begin{quote}{\bf Keywords:} #1\end{quote}}
\fi

\definecolor{grey}{cmyk}{0,0,0,0.7}

\definecolor{blue}{cmyk}{.6,0.1,0,0}
\definecolor{tblue}{rgb}{0,0,.7}
\definecolor{red}{cmyk}{0,.6,0.6,0}
\definecolor{green}{cmyk}{0.5,0,0.7,0}
\definecolor{magenta}{cmyk}{0,.8,0,0}
\definecolor{purple}{rgb}{.8,0,1}
\definecolor{lpurple}{cmyk}{.2,.6,0,0}
\definecolor{pink}{rgb}{1,.6,.6}
\definecolor{gold}{cmyk}{0,.1,.7,0}
\definecolor{orange}{cmyk}{0,.3,.6,0}
\definecolor{goldorange}{cmyk}{0,.2,.7,0}

\hypersetup{
pdfborder= 0 0 0,
colorlinks=true,
anchorcolor=grey,
citecolor=grey,
filecolor=grey,
linkcolor=grey,
menucolor=grey,
runcolor=grey,
urlcolor=grey
}

\renewcommand{\epsilon}{\varepsilon}
\renewcommand{\phi}{\varphi}

\newcommand{\NN}{{\mathbb N}}

\ifx\And\undefined
\newcommand{\And}{\wedge}
\fi

\ifx\AND\undefined
\newcommand{\AND}{\bigwedge}
\fi

\newcommand{\onlymath}{\null\ifmmode\else\errmessage{Your macro should only be used in math-mode}\fi}

\newcommand{\IGNORE}[1]{}

\input{psfig}

\newlength{\magic}

\makeatletter
\def\Picture[#1]{\@ifnextchar[{\@lPicture[#1]}{\@Picture[#1]}}
\def\@Picture[#1]#2{\par\begin{center}\@lPicture[#1][14]{#2}\end{center}\par}
\def\@lPicture[#1][#2]#3{\mbox{\vspace*{1ex}\hspace*{0.8cm}\psfig{rheight=#1cm,rwidth=#2cm,bbllx=3.1714cm,bblly=0cm,bburx=0cm,bbury=\magic,figure=#3.ps}\hspace*{-0.8cm}\vspace*{1ex}}}
\makeatother

\makeatletter
\def\Figure{\@ifnextchar[{\@lFigure}{\@Figure}}
\def\@Figure#1{\begin{center} \input{#1} \end{center}}
\def\@lFigure[#1]#2{\hspace*{#1}{\parbox{\textwidth}{\@Figure{#2}}}}
\makeatother

\definecolor{olivegreen}{cmyk}{.6,.2,0.6,0}

\usepackage{amsmath}

\usepackage{xspace}
\usepackage{listings}
\usepackage{cadp-exp}
\usepackage{cadp-lnt}
\usepackage{cadp-mcl}
\usepackage{cadp-svl}
\newcommand{\PIPE}{\textit{pipe}}

\title{
  Modeling an Asynchronous Circuit\\Dedicated to the Protection Against Physical Attacks
}
\def\titlerunning{Modeling an Asynchronous Circuit for Attack Protection}

\author{%
  Radu Mateescu \quad\qquad Wendelin Serwe
  \institute{Univ. Grenoble Alpes, Inria, CNRS, Grenoble INP\thanks{Institute of Engineering Univ. Grenoble Alpes},~~LIG, F-38000 Grenoble France}
  \email{First.Last@inria.fr}
  \\
  Aymane Bouzafour \quad\qquad Marc Renaudin
  \institute{Tiempo Secure -- SAS, Montbonnot-Saint-Martin, France}
  \email{First.Last@tiempo-secure.com}
}
\def\authorrunning{R.~Mateescu, W.~Serwe, A.~Bouzafour \& M.~Renaudin}

\begin{document}

\maketitle

\begin{abstract}
  Asynchronous circuits have several advantages for security applications, in particular their good resistance to attacks.
  In this paper, we report on experiments with modeling, at various abstraction levels, a patented asynchronous circuit for detecting physical attacks, such as cutting wires or producing short-circuits.
\end{abstract}

\section{Introduction}
\label{sec:introduction}

The number of connected devices that make up the IoT (\emph{Internet of Things}) already exceeded 20 billion, and is constantly increasing. However, a study of Hewlett Packard\footnote{\url{https://www8.hp.com/us/en/hp-news/press-release.html?id=1909050}} indicates that 90\% of the objects collect and thus potentially expose data, that 80\% of the objects do not use identification and source authentication mechanisms, and that 70\% do not use a mechanism of encrypting the transmitted data.
These vulnerabilities open the way to DDoS (\emph{Distributed Denial-of-Service}) attacks, which exploit over $100,000$ infected devices (e.g., cameras, video recorders, etc.) to overload various services and websites with a deluge of data.
Therefore, strengthening the security of connected devices is a critical issue.

The SECURIOT-2 project\footnote{\url{http://www.pole-scs.org/en/projects/securiot-2}}, led by Tiempo Secure, aims at developing a SMCU ({\em Secure Micro-Controller Unit\/}) that will bring to the IoT a level of security similar to banking transactions and transport (smart cards) and identification (electronic passports). Besides ensuring the necessary security services (key management, authentication, confidentiality and integrity of stored/exchanged data), the SMCU needs a power management scheme adequate with the low power consumption constraints of the IoT.

Given these constraints, a natural implementation of an SMCU is by means of asynchronous circuits, whose components are not governed by a clock signal, but operate independently, on an "on-demand" basis. Compared to classical synchronous circuits, this functioning has the potential for lower power dissipation, more harmonious electromagnetic emission, and better overall timing performance.

In this paper, we study the so-called {\em shield}~\cite{INPI-FR-3054344,EPO-EP-3276656-B1}, a particular asynchronous circuit designed and patented by Tiempo Secure for the protection of another (asynchronous) circuit against certain physical attacks (e.g., cutting a wire, setting a wire to a constant voltage, and introducing a short-circuit between two wires).
The behavior of asynchronous circuits can be suitably modeled in the action-based setting, using the composition operators of process calculi, as witnessed by CHP~\cite{Martin-86}, Tangram~\cite{Berkel-Kessels-Ronken-et-al-91}, Balsa~\cite{Edwards-Bardsley-02}, which are all inspired by CSP~\cite{Brookes-Hoare-Roscoe-84}, and have been successfully applied to design asynchronous circuits, e.g., \cite{Beigne-Clermidy-Vivet-Clouard-Renaudin-05,Martin-14,Plana-Riocreux-Bainbridge-et-al-03}.
All these approaches are equipped with a compilation scheme~\cite{Martin-86}, producing QDI ({\em Quasi-Delay-Insensitive\/}) circuits, the correct operation of which is independent of the delay in operators (or logical gates) and wires, except for certain wires that form {\em isochronic forks}~\cite{Martin-86}.
An isochronic fork is a wire connecting an emitter to several receivers, which receive any signal with identical delays.

We consider the modeling and analysis of the shield at two abstraction levels.
First, at circuit level, we take into account only the components of the circuit and their interconnection, without modeling the implementation details of these components. This level is appropriate for reasoning about the desired properties of the shield, namely the detection of physical attacks. The regular structure of the shield (serial pipeline) enables to apply inductive arguments to reduce all possible attack configurations to a finite set, which we analyzed exhaustively.
Next, we undertake a gate level modeling, focusing on the implementation of a component in terms of logical gates. Here we explore a range of different modeling variants for gates, electric wires, and forks (isochronic or not), and analyze their respective impact on the faithfulness of the global circuit model, the size of the underlying state spaces, the expression of correctness properties, and the overall ease of verification. We also point out that certain modeling variants lead to deadlocks in the circuit.

In this study, we mainly use the modern LNT~\cite{Garavel-Lang-Serwe-17} language for concurrent systems, which offers a user-friendly syntax and a formal semantics inherited from process calculi, and has been shown to be close to industrial hardware description languages for asynchronous circuits~\cite{Bouzafour-Renaudin-Garavel-et-al-18}.
For the modeling of attacks, we also use the synchronization vectors of EXP~\cite{Lang-05}, which operate on networks of communicating automata.
LNT and EXP are input languages of the CADP toolbox~\cite{Garavel-Lang-Mateescu-Serwe-13}\footnote{\url{http://cadp.inria.fr}} for modeling and verification of asynchronous concurrent systems.

The paper is organized as follows.
Section~\ref{sec:shield} describes the protection circuit and its behavior.
Sections~\ref{sec:circuit-level} and \ref{sec:gate-level} are devoted to the analysis of the protection circuit at two abstraction levels (circuit-level and gate-level, respectively), in regard to its properties of detecting physical attacks.
Both sections illustrate the approach on examples; the full models can be found in the appendices.
Section~\ref{sec:related} discusses related work and Section~\ref{sec:conclusion} gives concluding remarks and directions for future work.

\section{Shield}
\label{sec:shield}

To protect an integrated circuit against physical attacks, the patent~\cite{INPI-FR-3054344,EPO-EP-3276656-B1} suggests to add as a top layer a particular electric circuit, called \emph{shield} in the sequel.
Physical attacks to the circuit are generally meant to allow the attacker to probe for the sensitive information that is stored in the internal registers/wires of the circuit during its ordinary operation.
These probing procedures will inevitably lead to a tampering with the shield as it constitutes the top layer of the accessible part of the circuit.
Hence, proving that the shield is able to flag any tampering attempts amounts to proving that the circuit itself is able to detect a physical attack and to stop its normal operation and take an appropriate countermeasure (e.g., completely deactivate the circuit).

\begin{figure}[t]
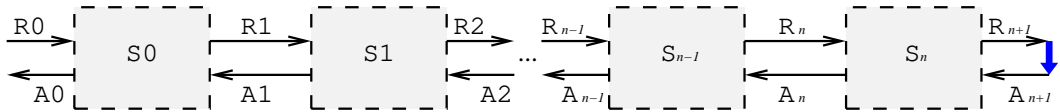

  \Figure{serial_shield_fig}
  \caption{Shield: serial pipelined composition of sequencers}
  \label{fig:shield}
\end{figure}

A shield is a serial composition of $n+1$ \emph{sequencers} (see Fig.~\ref{fig:shield} or \cite[Fig.~3]{EPO-EP-3276656-B1}), or even a parallel composition of several series of sequencers (see \cite[Figs.~9--11]{EPO-EP-3276656-B1}).
Each sequencer transmits a first signal, called \emph{request} to its successor; when the last sequencer outputs a request, a second signal, called \emph{acknowledgment} is transmitted through the series of sequencers in the opposite direction.
If a sequencer is designed in such a way that any physical modification on the connections for the transmission of the request (respectively, the acknowledgment) blocks the transmission of the acknowledgment (respectively, the request), a physical attack can be detected by the absence of an acknowledgment for a request sent into the shield.

\begin{figure}[t]
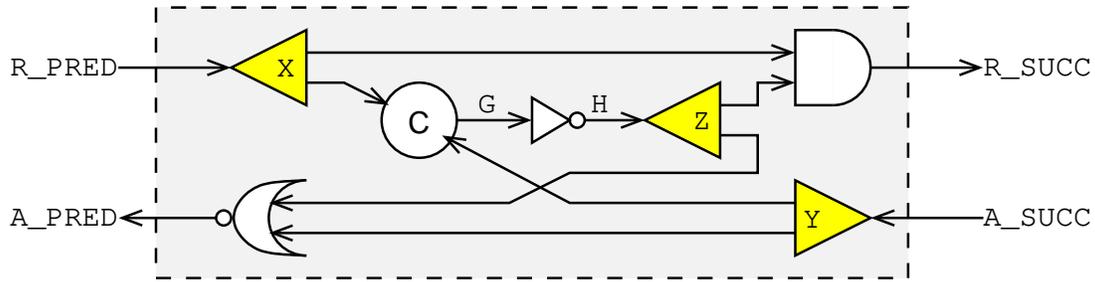

  \Figure{sequencer_fig}
  \caption{Gate-level design of a sequencer}
  \label{fig:sequencer}
\end{figure}

\begin{figure}[t]
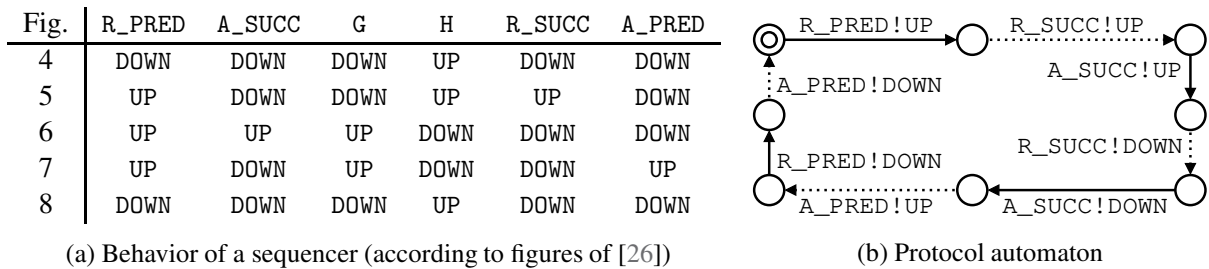

  \begin{minipage}{.60\textwidth}
  \begin{tabular}{c|cccccc}
    Fig. & \lstinline+R_PRED+ & \lstinline+A_SUCC+ & \lstinline+G+ & \lstinline+H+ & \lstinline+R_SUCC+ & \lstinline+A_PRED+ \\
    \hline
    4 & \lstinline+DOWN+ & \lstinline+DOWN+ & \lstinline+DOWN+ & \lstinline+UP+   & \lstinline+DOWN+ & \lstinline+DOWN+ \\
    5 & \lstinline+UP+   & \lstinline+DOWN+ & \lstinline+DOWN+ & \lstinline+UP+   & \lstinline+UP+   & \lstinline+DOWN+ \\
    6 & \lstinline+UP+   & \lstinline+UP+   & \lstinline+UP+   & \lstinline+DOWN+ & \lstinline+DOWN+ & \lstinline+DOWN+ \\
    7 & \lstinline+UP+   & \lstinline+DOWN+ & \lstinline+UP+   & \lstinline+DOWN+ & \lstinline+DOWN+ & \lstinline+UP+ \\
    8 & \lstinline+DOWN+ & \lstinline+DOWN+ & \lstinline+DOWN+ & \lstinline+UP+   & \lstinline+DOWN+ & \lstinline+DOWN+ \\
  \end{tabular}
  \subcaption{Behavior of a sequencer (according to figures of \cite{EPO-EP-3276656-B1})}
  \label{tab:sequencer}
  \end{minipage}
  ~
  \begin{minipage}{.38\textwidth}
    \Figure{protocol_cycle_fig}
    \vspace*{-1ex}
    \subcaption{Protocol automaton}
    \label{fig:protocol}
  \end{minipage}
  
  \caption{Behavior of a sequencer}
\end{figure}

Figure~\ref{fig:sequencer} shows the gate-level design of a circuit presented as a possible implementation of a sequencer in the patent.%
\footnote{%
  There are small differences with \cite[Figs.~4--8]{EPO-EP-3276656-B1}:
  (1) omission of the input \textsf{RST} to the Muller \textsf{C} element;
  (2) no distinction between links carrying \textsf{0} (dashed in~\cite{EPO-EP-3276656-B1}) and \textsf{1} (plain in~\cite{EPO-EP-3276656-B1});
  (3) naming forks and highlighting them (in yellow);
  (4) explicit drawing of the fork \lstinline+Z+ after the inverter; and
  (5) labeling of the visible and internal wires.
}
The intended evolution over time of the circuit~\cite[Figs.~4--8]{EPO-EP-3276656-B1} is summarized by the table in Fig.~\ref{tab:sequencer}.
When considering only the externally visible wires, the behavior of a sequencer corresponds to the automaton shown in Fig.~\ref{fig:protocol}, where input (respectively, output) transitions are depicted with plain (respectively, dotted) arrows.
Initially, the output \lstinline+G+ of the Muller \textsf{C} element~\cite{Muller-55} and the two inputs \lstinline+R_PRED+ and \lstinline+A_SUCC+ are \lstinline+DOWN+.
When input \lstinline+R_PRED+ goes \lstinline+UP+, output \lstinline+R_SUCC+ goes \lstinline+UP+ as well.
As a reaction, input \lstinline+A_SUCC+ is expected to go \lstinline+UP+, triggering output \lstinline+R_SUCC+ to go \lstinline+DOWN+.
When \lstinline+A_SUCC+ goes \lstinline+DOWN+, output \lstinline+A_PRED+ goes \lstinline+UP+.
Finally, \lstinline+R_PRED+ goes \lstinline+DOWN+, bringing the circuit back to its initial state.
This cycle is related to the so-called four-phase handshake protocol, frequently used in the design of asynchronous circuits~\cite{Martin-86}.
{}%
Note that there is no constraint on reaction delays of the circuit: a change of an input should be eventually followed by the corresponding output change.
Similarly, the environment of the circuit should respect the protocol by reacting to a change of an output by changing an input as specified by the protocol.

\section{Circuit-level Analysis of the Shield}
\label{sec:circuit-level}

\subsection{Modeling a Serial Shield}
\label{sec:serial-shield}

Considering a sequencer as a black-box, its behavior can be defined in the LNT modeling language~\cite{Garavel-Lang-Serwe-17} by the following process:
\begin{lstlisting}
process PROTOCOL [R_PRED, A_PRED, R_SUCC, A_SUCC: LINK] is
   loop
      R_PRED (UP);     R_SUCC (UP);   A_SUCC (UP);     R_SUCC (DOWN);
      A_SUCC (DOWN);   A_PRED (UP);   R_PRED (DOWN);   A_PRED (DOWN)
   end loop
end process
\end{lstlisting}
This process interacts with its environment on four LNT gates (\lstinline+R_PRED+, \lstinline+A_PRED+, \lstinline+R_SUCC+, and \lstinline+A_SUCC+), each of which is a channel carrying a voltage (\lstinline+DOWN+ or \lstinline+UP+).
The LTS (\emph{Labeled Transition System}) generated for process \lstinline+PROTOCOL+ is exactly the automaton shown in Figure~\ref{fig:protocol}.

Composition of sequencers in the manner of a pipeline as shown in Fig.~\ref{fig:shield} yields a model of a (serial) shield.
To ease the iterative construction of such sequence, we choose the EXP language~\cite{Lang-05} to define a composition operator $\PIPE(C_1, C_2)$ to pipe two sequencers $C_1$ and $C_2$ stored in the files \lstinline[language=exp]+"C1.bcg"+ and \lstinline[language=exp]+"C2.bcg"+:
\begin{lstlisting}[language=exp]
hide R, A in
   par R, A in
      rename R_SUCC -> R, A_SUCC -> A in "C1.bcg" end rename
   || rename R_PRED -> R, A_PRED -> A in "C2.bcg" end rename
   end par
\end{lstlisting}
Supposing that $C_1$ and $C_2$ both have the set \{\lstinline+R_PRED+, \lstinline+A_PRED+, \lstinline+R_SUCC+, \lstinline+A_SUCC+\} of observable gates, $\PIPE$ renames the right-hand (respectively, left-hand) side gates of $C_1$ (respectively, $C_2$) into a pair of new gates \{\lstinline+R+, \lstinline+A+\}, on which $C_1$ and $C_2$ are synchronized.
Hiding \lstinline+R+ and \lstinline+A+ in the composition ensures then that the observable gates of $\PIPE(C_1, C_2)$ are once more \{\lstinline+R_PRED+, \lstinline+A_PRED+, \lstinline+R_SUCC+, \lstinline+A_SUCC+\}.
Thus, $\PIPE$ can be easily extended to series of sequencers, by defining
$\PIPE^0 (S) = S$ and
$\PIPE^{n+1}(P) = \PIPE\bigl(P, \PIPE^n(P)\bigr)$.
{}

Letting $P$ stand for the LTS of process \lstinline+PROTOCOL+, equivalence checking (e.g., using BISIMULATOR~\cite{Bergamini-Descoubes-Joubert-Mateescu-05}) shows that $\PIPE(P, P) \equiv P$, where $\equiv$ denotes equivalence with respect to divergence-sensitive branching bisimulation. 
It follows by a straightforward induction that:
\begin{equation}
  \label{eq:induction}
  (\forall n \in \NN) \qquad \PIPE^n(P) \equiv P
\end{equation}

\subsection{Modeling Attacks}
\label{sec:circuit:attacks}

Intuitively, a physical attack on the shield corresponds to a modified composition of some sequencers, which can be represented by a modification of the composition operator.
The attacks can be grouped into two classes: those that impact the interface between two sequencers, and those that impact more than two sequencers.
If an attack can be expressed using operations for which divergence-sensitive branching bisimulation is a congruence, then it is sufficient to analyze this attack on any pair (or triple) of sequencers in a pipeline, because any non-trivial sequence of correctly pipelined sequencers can be reduced to a single sequencer.
In this section, we will present only one example of each class of attacks; a more complete treatment can be found in Appendix~\ref{app:exp}.

\subsubsection{Attacks impacting two sequencers: wire cuts and stuck at a constant}

Forcing a wire \lstinline+W+ to a constant value \lstinline+V+ can be represented by allowing only \lstinline+V+ as value during each rendezvous on the LNT gate \lstinline+W+.
Using the constraint-oriented specification style favored by the multi-way rendezvous~\cite{Brookes-Hoare-Roscoe-84,Hoare-85,Roscoe-Hoare-Bird-97,Garavel-Serwe-17}, it is sufficient to add a parallel process, which continuously accepts a rendezvous on \lstinline+W+ with value \lstinline+V+:
\begin{lstlisting}
process STUCKAT [W: LINK] (V: VOLTAGE) is
   loop
      W (V)
   end loop
end process
\end{lstlisting}
There are two possibilities to modify the composition operator: either both sequencers are synchronized with process \lstinline+STUCKAT+, or only the sequencer receiving on \lstinline+W+ (i.e., the right-hand side sequencer $C_2$ for gate \lstinline+R+, and the left-hand sequencer $C_1$ for gate \lstinline+A+).
The latter is expressed by the following composition expression in EXP, where file \lstinline+"STUCKAT_UP.bcg"+ contains the LTS of process \lstinline+STUCKAT [W] (UP)+:
\begin{lstlisting}[language=exp]
hide R, A in
   par
      A ->    rename R_SUCC -> R, A_SUCC -> A in "C1.bcg" end rename
   || R, A -> rename R_PRED -> R, A_PRED -> A in "C2.bcg" end rename
   || R ->    rename W -> R in "STUCKAT_UP.bcg" end rename
   end par
\end{lstlisting}
Note that \lstinline+"C1.bcg"+ can always perform a rendezvous on \lstinline+R+, without any constraint on the value.

A wire-cut can be handled similarly to a wire stuck at a constant: any rendezvous is blocked when synchronizing with the deadlocking process \lstinline+stop+ (rather than \lstinline+STUCKAT+), and any rendezvous is enabled by simply desynchronizing the two sequencers.

\subsubsection{Attacks impacting more than two sequencers: short-circuits}

\begin{figure}
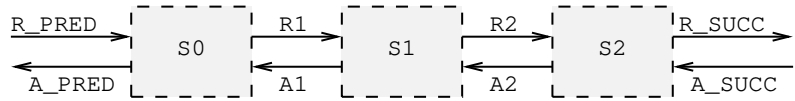

  \Figure{short_circuit_fig}
  \caption{Serial pipeline of three sequencers}
  \label{fig:short-circuits}
\end{figure}

For a series of three sequencers as shown in Fig.~\ref{fig:short-circuits}, there are six possible short-circuits around the sequencer \lstinline+S1+:
\lstinline+R1+-\lstinline+R2+, 
\lstinline+R1+-\lstinline+A1+, 
\lstinline+R2+-\lstinline+A1+, 
\lstinline+R2+-\lstinline+A2+, 
\lstinline+A1+-\lstinline+A2+, and 
\lstinline+A1+-\lstinline+R2+.%
\footnote{The short-circuits \lstinline+R1+-\lstinline+A1+ and \lstinline+R2+-\lstinline+A2+ are similar in the sense that they could both be defined by the same attack for a series of only two sequencers.}
The definition of a composition operator modeling a short-circuit requires the use of synchronization vectors, to properly handle any disagreement between the value exchanged during the rendezvous.
The most generic way to cope with these situations is to consider the resulting value to be non-deterministic.
Another, more constrained modeling option would be to enforce a majority-based policy (in a three-party rendezvous with two possible values, there is always a majority).

The following composition EXP expression models a short-circuit between \lstinline+R1+ and \lstinline+A2+:
\begin{lstlisting}[language=exp]
hide R1, A2, R2, A2, R1A2 in
  label par using
     -- synchronization vectors for unmodified wires
     "R_PRED !DOWN" * _          * _              -> "R_PRED !DOWN",
     "R_PRED !UP"   * _          * _              -> "R_PRED !UP",
     "A_PRED !DOWN" * _          * _              -> "A_PRED !DOWN",
     "A_PRED !UP"   * _          * _              -> "A_PRED !UP",
     "A1 !DOWN"     * "A1 !DOWN" * _              -> "A1 !DOWN",
     "A1 !UP"       * "A1 !UP"   * _              -> "A1 !UP",
     _              * "R2 !DOWN" * "R2 !DOWN"     -> "R2 !DOWN",
     _              * "R2 !UP"   * "R2 !UP"       -> "R2 !UP",
     _              * _          * "R_SUCC !DOWN" -> "R_SUCC !DOWN",
     _              * _          * "R_SUCC !UP"   -> "R_SUCC !UP",
     _              * _          * "A_SUCC !DOWN" -> "A_SUCC !DOWN",
     _              * _          * "A_SUCC !UP"   -> "A_SUCC !UP",
     -- synchronization vectors for the short-circuit
     "R1 !DOWN"     * "R1 !DOWN" * "A2 !DOWN"     -> "R1A2 !DOWN",
     "R1 !DOWN"     * "R1 !DOWN" * "A2 !UP"       -> "R1A2 !DOWN",
     "R1 !DOWN"     * "R1 !DOWN" * "A2 !UP"       -> "R1A2 !UP",
     "R1 !UP"       * "R1 !UP"   * "A2 !DOWN"     -> "R1A2 !DOWN",
     "R1 !UP"       * "R1 !UP"   * "A2 !DOWN"     -> "R1A2 !UP",
     "R1 !UP"       * "R1 !UP"   * "A2 !UP"       -> "R1A2 !UP",
     "R1 !DOWN"     * "A2 !DOWN" * "A2 !DOWN"     -> "R1A2 !DOWN",
     "R1 !DOWN"     * "A2 !UP"   * "A2 !UP"       -> "R1A2 !DOWN",
     "R1 !DOWN"     * "A2 !UP"   * "A2 !UP"       -> "R1A2 !UP",
     "R1 !UP"       * "A2 !DOWN" * "A2 !DOWN"     -> "R1A2 !DOWN",
     "R1 !UP"       * "A2 !DOWN" * "A2 !DOWN"     -> "R1A2 !UP",
     "R1 !UP"       * "A2 !UP"   * "A2 !UP"       -> "R1A2 !UP"
  in
     rename R_SUCC -> R1, A_SUCC -> A1 in "C1.bcg" end rename
  || rename R_PRED -> R1, A_PRED -> A1, R_SUCC -> R2, A_SUCC -> A2 in
        "C2.bcg"
     end rename
  || rename R_PRED -> R2, A_PRED -> A2 in "C3.bcg" end rename
  end par
\end{lstlisting}
There are two kinds of rules for unmodified wires: there is no synchronization for those externally visible (\lstinline+R_PRED+, \lstinline+A_PRED+, \lstinline+R_SUCC+, and \lstinline+A_SUCC+), and a binary synchronization for the remaining ones (\lstinline+A1+ and \lstinline+R2+).
For the three-party rendezvous of the short-circuit (in the example between \lstinline+R1+ and \lstinline+A2+), a new gate \lstinline+R1A2+ is introduced; the voltage on \lstinline+R1A2+ is non-deterministic if and only if there is disagreement on the voltage.

\subsubsection{Circuit-Level Analysis Results}

The SVL (\emph{Script Verification Language})~\cite{Garavel-Lang-01} script given in Appendix~\ref{app:svl} automates the circuit-level validation of the shield.
To show that the shield detects an attack, we check whether the expected behavior (see Fig.~\ref{fig:protocol}) is included in the model generated for the corresponding attack.

All attacks forcing a wire to a constant value are detected.
A wire-cut might be left undetected only if the attack is modeled such that the gates on both ends of the wire are free to perform arbitrary rendezvous on the gate without ever synchronizing, which is not a realistic assumption.
All short-circuits but \lstinline+R1+-\lstinline+A1+ and \lstinline+R2+-\lstinline+A2+ are detected.
This can be explained by the fact that these short-circuits cut the pipe-line of sequencers, reducing them to a shorter shield, keeping the same functionality.
However, because the electric connections for \lstinline+R+ and \lstinline+A+ are located in different layers of the chip~\cite{INPI-FR-3054344,EPO-EP-3276656-B1}, attacking the chip in this way is impossible (without damaging the entire chip).

Because attacks are modeled by operators congruent for branching bisimulation, these results extend to a series of $N$ sequencers.
Indeed, using equation~\eqref{eq:induction} any non-trivial series of sequencers can be replaced by a single sequencer, so that any (single) attack can be reduced to one of the studied configurations with two or three sequencers.

{}

{}

\section{Gate-level Analysis}
\label{sec:gate-level}

For a more precise study of a sequencer, we refine the model of a sequencer by representing the internal operation of all the electric wires and gates according to Fig.~\ref{fig:sequencer}: one binary wire, three forks, one inverter, one AND gate, one NOR gate, and one Muller \textsf{C} element.
Intuitively, each of them can be considered as a process, which reacts on its input(s) by (possibly) producing some output.
However, there are various possibilities with respect to the degree of synchronization and possible propagation delays.
This section aims at exploring these possibilities, discussing the respective advantages in terms of ease of modeling, size of the corresponding state space, and practicality for verification.

A major choice when modeling a gate is whether the model represents only \emph{transitions} (i.e., changing voltage, as in the protocol of Fig.~\ref{fig:protocol}) or rather current \emph{state} of a wire.
Considering only transitions yields smaller models and corresponding state spaces, but might seem less intuitive because it requires stronger assumptions (e.g., synchronous communication).

{}

\subsection{Modeling Wires and Forks}

A wire can be modeled in essentially two ways: as an LNT gate or as an instance of the LNT process
\begin{lstlisting}
process WIRE [INPUT, OUTPUT: LINK] is
   var X : VOLTAGE in
      loop
         INPUT (?X);  OUTPUT (X)
      end loop
   end var
end process
\end{lstlisting}
Representing a wire by an LNT gate models the immediate transmission from the input to the output, whereas the LNT process models the possibility of a communication delay, because other actions of the circuit might be interleaved between the input and the output.
Notice that (similar to \cite{Kapoor-Josephs-04}) process \lstinline+WIRE+ enforces alternation, i.e., it only accepts the next input after it has delivered the output.
{}

A fork is a wire with more than one output, and can also be modeled as a gate (using a multi-way rendezvous~\cite{Brookes-Hoare-Roscoe-84,Hoare-85,Roscoe-Hoare-Bird-97,Garavel-Serwe-17}), faithfully representing isochronic forks~\cite{Josephs-07}.
It is also possible to model a fork as a dedicated process.
However, as there is more than one output, it is possible to specify their order of occurrence, from simultaneous (using a multi-way rendezvous for the outputs) to unspecified (using a parallel composition).
The latter option is modeled by the LNT process
\begin{lstlisting}
process FORK [INPUT, OUTPUT1, OUTPUT2: LINK] is
   var X : VOLTAGE in
      loop
         INPUT (?X);
         par
            OUTPUT1 (X)
         || OUTPUT2 (X)
         end par
      end loop
   end var
end process
\end{lstlisting}

To ensure a correct functioning of an asynchronous circuit, some forks must be assumed to be \emph{isochronic}.
An isochronic fork can be modeled using the process \lstinline+WIRE+, using a multi-way rendezvous on gate \lstinline+OUTPUT+ (similar to~\cite{Josephs-07}).
{}

Notice that all these models of wires and forks are valid for both the transition and state oriented style, because they do not have a memory about the last transmitted value.

\subsection{Modeling a Sequencer}

Supposing that we have LNT processes modeling gates, the sequencer of Fig.~\ref{fig:sequencer} can be modeled by a spectrum of possibilities, considering the options of modeling wires and forks.
A first model using multi-way rendezvous for wires and forks is:
\begin{lstlisting}
process SEQUENCER_RV [R_PRED, A_PRED, R_SUCC, A_SUCC, G, H: LINK]
                     (X1, X2, INIT_C: VOLTAGE) is
   par
      R_PRED, A_SUCC, G -> MULLER [R_PRED, A_SUCC, G] (X1, X2, INIT_C)
   || R_PRED, H -> AND [R_PRED, H, R_SUCC] (X1, NOT (INIT_C))
   || G, H -> INV [G, H] (INIT_C)
   || A_SUCC, H -> NOR [A_SUCC, H, A_PRED] (X2, NOT (INIT_C))
   end par
end process
\end{lstlisting}

On the other extreme, the following LNT process represents wires and forks as explicit processes, and all its forks but \lstinline+Z+ are isochronic:
\begin{lstlisting}
process SEQUENCER_IIP [R_PRED, A_PRED, R_SUCC, A_SUCC, G, H,
                       R_PRED2, A_SUCC2, G2, H1, H2: LINK]
                      (X1, X2, INIT_C: VOLTAGE) is
   par
      R_PRED2, A_SUCC2, G ->
         MULLER [R_PRED2, A_SUCC2, G] (X1, X2, INIT_C)
   || R_PRED2, H1 -> AND [R_PRED2, H1, R_SUCC] (X1, NOT (INIT_C))
   || G2, H -> INV [G2, H] (INIT_C)
   || A_SUCC2, H2 -> NOR [A_SUCC2, H2, A_PRED] (X2, NOT (INIT_C))
   || R_PRED2 -> WIRE [R_PRED, R_PRED2]
   || A_SUCC2 -> WIRE [A_SUCC, A_SUCC2]
   || H, H1, H2 -> FORK [H, H1, H2]
   || G, G2 -> WIRE [G, G2]
   end par
end process
\end{lstlisting}
Models for other combinations are given in Appendix~\ref{app:lnt:sequencer}.

\subsection{Modeling Gates}
\label{sec:gate:and}

\begin{figure}
  \begin{tabular}{@{}ccc@{}}
    \begin{minipage}[t]{.34\textwidth}
\begin{lstlisting}
loop
   select
      INPUT1 (?X1)
   [] INPUT2 (?X2)
   end select;
   OUTPUT (X1 AND X2)
end loop
\end{lstlisting}

      \centering
      (a) intuitive

\begin{lstlisting}
loop
   select
      INPUT1 (?X1)
   [] INPUT2 (?X2)
   [] OUTPUT (X1 AND X2)
   end select
end loop
\end{lstlisting}

      \centering
      (d) free
    \end{minipage}
    &
    \begin{minipage}[t]{.30\textwidth}
\begin{lstlisting}
loop
   select
      INPUT1 (?X1);
      select
         null
      [] INPUT2 (?X2)
      end select
   [] INPUT2 (?X2);
      select
         null
      [] INPUT1 (?X1)
      end select
   end select;
   OUTPUT (X1 AND X2)
end loop
\end{lstlisting}

      \centering
      (b) state-oriented
    \end{minipage}
    &
    \begin{minipage}[t]{.30\textwidth}
\begin{lstlisting}
loop
   par
      select
         INPUT1 (?X1)
      [] null
      end select 
   || select
         INPUT2 (?X2)
      [] null
      end select
   end par;
   OUTPUT (X1 AND X2)
end loop
\end{lstlisting}

      \centering
      (c) parallel
    \end{minipage}
  \end{tabular}
  \caption{State-oriented models of a binary AND gate}
  \label{fig:and}
\end{figure}

We illustrate the various ways to model gates on the binary \lstinline+AND+ gate; models of the other gates are given in Appendix~\ref{app:lnt}.
Fig.~\ref{fig:and} shows various (state-oriented) bodies that can replace ``\lstinline+...+'' in the LNT process
\begin{lstlisting}
process AND [INPUT1, INPUT2, OUTPUT: LINK] (in var X1, X2: VOLTAGE) is
   ...
end process
\end{lstlisting}
The most intuitive model is state-oriented and shown in Fig.~\ref{fig:and}(a).
It reacts to a rendezvous on one of its inputs with a rendezvous on its output.
An issue with Fig.~\ref{fig:and}(a) is that it requires an output to occur between any two inputs, but it might be the case that the inputs arrive almost simultaneously.
This issue is addressed by Fig.~\ref{fig:and}(b), which may accept one or two inputs in an arbitrary order before generating an output.
Using a parallel composition operator, Fig.~\ref{fig:and}(b) can be simplified to Fig.~\ref{fig:and}(c).
Fig.~\ref{fig:and}(c) has the inconvenient that it might generate outputs that are not triggered by an input.
Permitting also that an input might not generate an output, one obtains Fig.~\ref{fig:and}(d), which is always ready to accept a new input or to (re)generate the current output.

A transition-oriented model of a gate must generate an output in reaction to a change in one of the inputs if and only if the output changes.
This requires to remember the previous output, as in the process:
\begin{lstlisting}
process AND [INPUT1, INPUT2, OUTPUT: LINK] (in var X1, X2: VOLTAGE) is
   var RESULT, NEW_RESULT: VOLTAGE in
      RESULT := X1 AND X2;
      loop
         select
            INPUT1 (?X1)
         [] INPUT2 (?X2)
         end select;
         NEW_RESULT := X1 AND X2; 
         if NEW_RESULT != RESULT then
            RESULT := NEW_RESULT;
            OUTPUT (RESULT)
         end if
      end loop
   end var
end process
\end{lstlisting}

\subsection{Gate-level Analysis Results}

\begin{table}
  \small\centering
\begin{tabular}{@{}c@{~~}crrrrcrrc@{}}
  model & forks &
  \multicolumn{2}{c}{one sequencer} &
  \multicolumn{3}{c}{one sequencer with stubs} &
  \multicolumn{3}{c}{two sequencers}
  \\
  & &
  \multicolumn{1}{c}{states} & \multicolumn{1}{c}{transitions} &
  \multicolumn{1}{c}{states} & \multicolumn{1}{c}{transitions} &
  lock &
  \multicolumn{1}{c}{states} & \multicolumn{1}{c}{transitions} &
  lock
  \\
  \hline\hline
  & RV  &      90 &       222 &       8 &         8 & no  &           308 &           790 & yes \\
  & III &   2,586 &     7,922 &       8 &         8 & no  &       288,771 &     1,093,552 & yes \\
  & IIP &   6,124 &    21,454 &       8 &         8 & no  &     1,307,889 &     5,968,266 & yes \\
  & IPI &   6,475 &    19,985 &   1,444 &     4,141 & yes &     2,588,785 &    10,624,729 & yes \\
  \rotatebox{90}{\makebox[1em]{INTUITIVE}}
  & IPP &  15,422 &    54,969 &   4,780 &    15,402 & yes &    12,433,518 &     61,288,551 & yes \\
  & PII &   6,475 &    19,985 &   1,444 &     4,141 & yes &     1,562,907 &      6,452,580 & yes \\
  & PIP &  15,422 &    54,969 &   4,780 &    15,402 & yes &     7,291,527 &     36,213,052 & yes \\
  & PPI &  14,900 &    46,486 &   4,032 &    13,710 & yes &    13,450,533 &     59,014,624 & yes \\
  & PPP &  38,680 &   139,558 &  14,273 &    53,496 & yes &    76,518,596 &    400,442,323 & yes \\
  \hline
  & RV  &     766 &     2,406 &       8 &         8 & no  &       230,906 &        906,342 & no \\
  & III &  33,258 &   127,380 &       8 &         8 & no  &   474,187,601 &  2,514,512,879 & no \\
  & IIP &  86,846 &   374,292 &       8 &         8 & no  & 3,002,896,049 & 18,494,246,894 & no \\
  & IPI &  82,041 &   315,312 &  32,990 &   113,773 & no  & 2,780,162,577 & 15,419,740,546 & no \\ 
  \rotatebox{90}{\makebox[1em]{STATE}}
  & IPP & 216,470 &   931,696 &  89,238 &   339,090 & no  & \\ 
  & PII &  82,041 &   315,312 &  32,990 &   113,773 & no  & 2,795,890,977 & 15,509,939,437 & no \\ 
  & PIP & 216,470 &   931,696 &  89,238 &   339,090 & no  & \\ 
  & PPI & 194,738 &   752,128 &  82,020 &   316,170 & no  & \\ 
  & PPP & 531,576 & 2,287,840 & 238,574 & 1,000,138 & no  & \\ 
  \hline
  & RV  &     916 &     3,404 &       8 &        16 & no  &       342,674 &      1,625,792 & no \\
  & III &  54,394 &   258,456 &       8 &        16 & no  & 1,308,613,124 &  8,868,967,479 & no \\
  & IIP & 146,002 &   734,800 &       8 &        16 & no  & 8,464,022,990 & 61,740,118,299 & no \\
  & IPI & 127,578 &   606,064 &  72,431 &   301,189 & no  & 6,962,294,015 & 48,364,449,041 & no \\ 
  \rotatebox{90}{\makebox[1em]{PARALLEL}}
  & IPP & 339,047 & 1,704,667 & 194,704 &   854,767 & no  & \\
  & PII & 127,578 &   606,064 &  72,431 &   301,189 & no  & 7,000,306,907 & 48,656,915,179 & no \\
  & PIP & 339,047 & 1,704,667 & 194,704 &   854,767 & no  & \\
  & PPI & 292,906 & 1,391,688 & 173,252 & 1,115,840 & no  & \\
  & PPP & 778,468 & 3,913,592 & 474,676 & 2,254,762 & no  & \\
  \hline
  & RV  &      24 &       186 &       8 &        16 & no  &         567 &          6,517 & no \\
  & III &     384 &     2,664 &       8 &        16 & no  &     147,360 &      1,602,532 & no \\
  & IIP &     768 &     5,544 &       8 &        16 & no  &     589,440 &      6,741,584 & no \\
  & IPI &     764 &     5,306 &   7,145 &    37,733 & no  &     583,560 &      6,500,110 & no \\
  \rotatebox{90}{\makebox[1em]{FREE}}
  & IPP &   1,528 &    11,040 &  14,346 &    80,865 & no  &   2,334,240 &     27,307,944 & no \\
  & PII &     764 &     5,306 &   7,145 &    37,733 & no  &     582,224 &      6,485,364 & no \\
  & PIP &   1,528 &    11,040 &  14,346 &    80,865 & no  &   2,328,896 &     27,245,680 & no \\
  & PPI &   1,520 &    10,568 &  15,764 &    90,732 & no  &   2,310,400 &     26,344,640 & no \\
  & PPP &   3,040 &    21,984 &  33,774 &   206,894 & no  &   9,241,600 &    110,534,400 & no \\
  \hline
  & RV  &      34 &       112 &       8 &         8 & no  &         279 &          1,101 & no \\
  & III &     496 &     1,614 &       8 &         8 & no  &      34,461 &        153,267 & no \\
  & IIP &   1,320 &     4,870 &       8 &         8 & no  &     238,811 &      1,270,154 & no \\
  & IPI &     952 &     3,155 &     702 &     2,077 & yes &     136,092 &        665,059 & yes \\
  \rotatebox{90}{\makebox[1em]{TRANSITION}}
  & IPP &   2,475 &     9,313 &   1,938 &     6,525 & yes &     912,702 &      5,269,597 & yes \\
  & PII &     952 &     3,155 &     702 &     2,077 & yes &     135,814 &        666,185 & yes \\
  & PIP &   2,475 &     9,313 &   1,938 &     6,525 & yes &     911,332 &      5,263,119 & yes \\
  & PPI &   1,814 &     6,104 &   1,335 &     8,233 & yes &     537,434 &      2,855,374 & yes \\
  & PPP &   4,712 &    17,972 &   4,789 &    18,942 & yes &   3,624,160 &     22,403,699 & yes \\
\end{tabular}
  \caption{State spaces for various gate-level models of up to two sequencers}
  \label{tab:results}
\end{table}

We used the Grid'5000 platform to generate the state spaces for one and two sequencers, using various combinations of the modeling choices discussed previously.
Table~\ref{tab:results} summarizes the results.
The first column indicates the model of gates: INTUITIVE, STATE, PARALLEL, and FREE refer respectively to the state-oriented style of Fig.~\ref{fig:and}(a), (b), (c), and (d), and TRANSITION refers to the transition-oriented style (see Sect.~\ref{sec:gate:and}).
The second column indicates the model of wires and forks: RV indicates that wires and forks are modeled by a rendezvous, and a triple $F_1F_2F_3$ (with $F_i \in \{\text{I}, \text{P}\}$) indicates the isochrony of the three forks (\lstinline+X+, \lstinline+Y+, and \lstinline+Z+ in Fig.~\ref{fig:sequencer}) in a sequencer (I stands for an isochronic fork, and P for a non-isochronic one); in all cases but for RV, wires are modeled as separate processes.
The third and fourth columns indicate the size of one sequencer, minimized for divergence-sensitive branching bisimulation (after hiding all gates but \lstinline+R_PRED+, \lstinline+A_PRED+, \lstinline+R_SUCC+, and \lstinline+A_SUCC+).
The fifth and sixth columns indicate the size of one sequencer with stubs (generating input according to the protocol and absorbing repeated outputs, see Appendix~\ref{app:lnt:stubs}), minimized for divergence-sensitive branching bisimulation.
The seventh column indicates whether a sequencer with stubs contains a deadlock (after minimizing for branching bisimulation to remove livelocks masking deadlocks).
The eighth and nineth columns indicate the size of a pipeline of two sequencers (not minimized for branching bisimulation), whenever the composition was possible.
The last column indicates whether the model of two sequencers contains a deadlock.
The most time-consuming task is the (explicit) generation of the LTS for the composition of two sequencers: depending on the model, this step takes from five seconds to more than several days.%
\footnote{%
  For the missing results for models STATE and PARALLEL, the generation was stopped after 62 hours, at which point each composition had required already 150 GB of RAM and produced a file of more than 100 GB.
  We also experimented with distributed generation, using up to 80 processors; this also failed due to a lack of disk space (more than 4 TB).
  {}
}

{}

A first observation is that explicitly modeling wires and forks increases the size of the models by several orders of magnitude (compared to models RV).
Also, the state-oriented (other than FREE) models are significantly larger then the transition-oriented one.

A second observation is the presence of deadlocks in some compositions of two sequencers.
For instance, in the INTUITIVE model with wires and forks represented by LNT gates using rendezvous (RV), the following sequence of rendezvous leads to the deadlock:
\lstinline+"R_PRED !UP"+,
\lstinline+"R !UP"+,
\lstinline+"R_SUCC !UP"+,
\lstinline+"G_L !DOWN"+,
\lstinline+"G_R !DOWN"+,
\lstinline+"A_SUCC !UP"+,
\lstinline+"R_PRED !UP"+.
The deadlock can be explained by the fact that the composition of two sequencers contains a cycle (formed by the (forking) wires \lstinline+R+, \lstinline+G_R+, \lstinline+H_R+, \lstinline+A+, \lstinline+G_L+, \lstinline+H_L+)\footnote{See Appendix~\ref{app:exp:pipe} for the corresponding composition expression.}, which can absorb only a bounded number of inputs (\lstinline+R_PRED+ and \lstinline+A_SUCC+).
Because each input must be absorbed twice by the cycle (e.g., \lstinline+R_PRED+ has to alternate with both \lstinline+R+ and \lstinline+G_L+ and in the trace above), the cycle can fill if inputs arrive faster than outputs (\lstinline+R_SUCC+ and \lstinline+A_PRED+) are generated.
{}
{}
We deduce that the INTUITIVE model is not appropriate.
We also observe that the presence of deadlocks in the TRANSITION models can be used to pinpoint the forks that need not to be isochronic; this provides a valuable information to the hardware designer about possible substantial optimizations in area and performance.

{}

A third observation is that, contrary to the circuit-level model, the composition of two sequencers in the gate-level models is not equivalent to a single sequencer, so that there is no counterpart to the induction result of Section~\ref{sec:circuit-level}.
However, when constraining the visible actions (\lstinline+R_PRED+, \lstinline+A_PRED+, \lstinline+R_SUCC+, and \lstinline+A_SUCC+) by appropriate stubs eliminating repeated actions (see Appendix~\ref{app:lnt:stubs}), we observe that a sequencer is branching bisimilar to the protocol if and only if the forks \lstinline+X+ and \lstinline+Y+ are both isochronic (i.e., models RV, III, and IIP).%
\footnote{%
  Table~\ref{tab:results} indicates the size of the sequencers with stubs reduced for divergence-sensitive branching bisimulation.
  Because the models PARALLEL and FREE contain livelocks (as a gate might generate outputs not triggered by inputs), there are 16 transitions for RV, III, and IIP (one livelock per state).
}
Hence, the LNT model is useful to determine whether a fork must be isochronic (this is the case for \lstinline+X+ and \lstinline+Y+ for the sequencer) or not (this is the case for \lstinline+Z+, and formally justifies why \lstinline+Z+ is represented differently than \lstinline+X+ and \lstinline+Y+ in \cite[Figs.~4--8]{EPO-EP-3276656-B1}.

Theoretically, attacks could be analyzed as described in Sect.~\ref{sec:circuit:attacks} for circuit-level models.
However, because the state spaces of gate-level models are much larger, we did not yet attempt this{}.

\section{Related Work}
\label{sec:related}

Process calculi, that were initially designed for concurrent systems, are natural candidates for describing asynchronous circuits.

CSP$_M$, a machine-readable dialect of CSP, was used in~\cite{Josephs-07} to model various asynchronous circuits (C element, $n$-way mutual exclusion element, tree arbiter) and verify them using the FDR tool. The CSP operators were shown to be suitable for asynchronous circuits, in particular multi-way synchronization facilitates the modeling of isochronic forks.
A general model for asynchronous circuits and its translation to CSP is proposed in~\cite{Wang-Kwiatkowska-07}, together with a compositional verification approach suitable for FDR. This makes possible a modeling at three levels (Balsa, handshake expansion, and gate-level), and was applied to the design of realistic circuits, such as the AMULET processor~\cite{Wang-Kwiatkowska-et-al-06}.

A modeling style for asynchronous circuits in CCS was proposed in~\cite{Stevens-Aldwinckle-Birtwistle-Liu-93} and illustrated on distributed arbiters.
In~\cite{Kapoor-Josephs-04} it is shown how to represent DI ({\em Delay-Insensitive\/}) asynchronous circuits in CCS. A circuit $C$ was defined as DI if its composition with a FRW (\emph{Foam Rubber Wrapper}) making the communications on input and output wires arbitrarily long is equivalent to $C$ modulo the MUST-testing equivalence, provided by the Concurrency Workbench~\cite{Cleaveland-Parrow-Steffen-89}.

The modeling and verification (by equivalence checking) of asynchronous designs using Circal is discussed in~\cite{Bailey-McCaskill-Milne-94}. It is also shown how the diagnostic facility of the Circal system helps in determining the forks that are required to be isochronic.

Besides process calculi, a number of other formalisms have been used to model asynchronous circuits at gate-level:
Petri nets~\cite{Yakovlev-Koelmans-Semenov-et-al-96},
signal transition graphs~\cite{Yakovlev-Kishinevsky-Kondratyev-et-al-96},
XDI~\cite{Verhoeff-98},
Receptive Process Theory~\cite{Josephs-92},
stable events~\cite{Jia-Li-He-17},
and complete trace structures~\cite{Dill-88}.

{}

We do not consider here the verification of security properties, because the main focus of the present paper is the faithful modeling of an asynchronous circuit regardless of its purpose.

\section{Conclusion}
\label{sec:conclusion}

In this paper, we used the LNT language to formally model an asynchronous circuit at circuit- and gate-level, investigating also various modeling styles for wires and gates.
Analyzing these models using the CADP toolbox, we found that they can provide the designer with valuable information, such as the necessity to ensure isochrony of forks.
For circuit-level analysis, we used an induction proof to extend results of our attack analysis to a shield of arbitrary size.
Obtaining a comparable result for gate-level models is challenging, due to the necessity of stubs.
Last, but not least, we believe that, due to their extensibility and state space size, these models provide a challenging benchmark.

\paragraph*{Acknowledgments.}

The present work has been partly funded by BPI France and FEDER (\emph{Fonds Européen de Développement Économique Régional}) Rhône-Alpes Auvergne under the national project ``SecurIoT-2'' supported by the four competitiveness clusters Minalogic, SCS, Systematic Paris-Région, and Derbi.
Experiments presented in this paper were carried out using the Grid'5000 testbed, supported by a scientific interest group hosted by Inria and including CNRS, RENATER and several Universities as well as other organizations (see \url{https://www.grid5000.fr}).

\bibliographystyle{eptcs}
\bibliography{shield}

\begin{thebibliography}{10}
\providecommand{\bibitemdeclare}[2]{}
\providecommand{\surnamestart}{}
\providecommand{\surnameend}{}
\providecommand{\urlprefix}{Available at }
\providecommand{\url}[1]{\texttt{#1}}
\providecommand{\href}[2]{\texttt{#2}}
\providecommand{\urlalt}[2]{\href{#1}{#2}}
\providecommand{\doi}[1]{doi:\urlalt{http://dx.doi.org/#1}{#1}}
\providecommand{\bibinfo}[2]{#2}

\bibitemdeclare{article}{Bailey-McCaskill-Milne-94}
\bibitem{Bailey-McCaskill-Milne-94}
\bibinfo{author}{Andrew \surnamestart Bailey\surnameend},
  \bibinfo{author}{George~A. \surnamestart McCaskill\surnameend} \&
  \bibinfo{author}{George~J. \surnamestart Milne\surnameend}
  (\bibinfo{year}{1994}): \emph{\bibinfo{title}{An exercise in the automatic
  verification of asynchronous designs}}.
\newblock {\sl \bibinfo{journal}{Formal Methods in System Design}}
  \bibinfo{volume}{4}(\bibinfo{number}{3}), pp. \bibinfo{pages}{213--242},
  \doi{10.1007/BF01384047}.

\bibitemdeclare{inproceedings}{Beigne-Clermidy-Vivet-Clouard-Renaudin-05}
\bibitem{Beigne-Clermidy-Vivet-Clouard-Renaudin-05}
\bibinfo{author}{Edith \surnamestart Beign\'e\surnameend},
  \bibinfo{author}{Fabien \surnamestart Clermidy\surnameend},
  \bibinfo{author}{Pascal \surnamestart Vivet\surnameend},
  \bibinfo{author}{Alain \surnamestart Clouard\surnameend} \&
  \bibinfo{author}{Marc \surnamestart Renaudin\surnameend}
  (\bibinfo{year}{2005}): \emph{\bibinfo{title}{An Asynchronous {NoC}
  Architecture Providing Low Latency Service and Its Multi-Level Design
  Framework}}.
\newblock In: {\sl \bibinfo{booktitle}{Proceedings of the 11th IEEE
  International Symposium on Asynchronous Circuits and Systems ASYNC'05 (New
  York, USA)}}, \bibinfo{publisher}{IEEE Computer Society Press}, pp.
  \bibinfo{pages}{54--63}, \doi{10.1109/ASYNC.2005.10}.

\bibitemdeclare{inproceedings}{Bergamini-Descoubes-Joubert-Mateescu-05}
\bibitem{Bergamini-Descoubes-Joubert-Mateescu-05}
\bibinfo{author}{Damien \surnamestart Bergamini\surnameend},
  \bibinfo{author}{Nicolas \surnamestart Descoubes\surnameend},
  \bibinfo{author}{Christophe \surnamestart Joubert\surnameend} \&
  \bibinfo{author}{Radu \surnamestart Mateescu\surnameend}
  (\bibinfo{year}{2005}): \emph{\bibinfo{title}{BISIMULATOR: A Modular Tool for
  On-the-Fly Equivalence Checking}}.
\newblock In \bibinfo{editor}{Nicolas \surnamestart Halbwachs\surnameend} \&
  \bibinfo{editor}{Lenore \surnamestart Zuck\surnameend}, editors: {\sl
  \bibinfo{booktitle}{Proceedings of the 11th International Conference on Tools
  and Algorithms for the Construction and Analysis of Systems (TACAS'05),
  Edinburgh, Scotland, UK}}, {\sl \bibinfo{series}{Lecture Notes in Computer
  Science}} \bibinfo{volume}{3440}, \bibinfo{publisher}{Springer Verlag}, pp.
  \bibinfo{pages}{581--585}, \doi{10.1007/978-3-540-31980-1_42}.

\bibitemdeclare{inproceedings}{Berkel-Kessels-Ronken-et-al-91}
\bibitem{Berkel-Kessels-Ronken-et-al-91}
\bibinfo{author}{Kees \surnamestart van Berkel\surnameend},
  \bibinfo{author}{Joep \surnamestart Kessels\surnameend},
  \bibinfo{author}{Marly \surnamestart Roncken\surnameend},
  \bibinfo{author}{Ronald \surnamestart Saeijs\surnameend} \&
  \bibinfo{author}{Frits \surnamestart Schalij\surnameend}
  (\bibinfo{year}{1991}): \emph{\bibinfo{title}{The {VLSI}-Programming Language
  Tangram and its Translation into Handshake Circuits}}.
\newblock In: {\sl \bibinfo{booktitle}{Proceedings of the Conference on
  European Design Automation (Amsterdam, The Netherlands)}},
  \bibinfo{publisher}{IEEE Computer Society Press}, pp.
  \bibinfo{pages}{384--389}, \doi{10.1109/EDAC.1991.206431}.

\bibitemdeclare{inproceedings}{Bouzafour-Renaudin-Garavel-et-al-18}
\bibitem{Bouzafour-Renaudin-Garavel-et-al-18}
\bibinfo{author}{Aymane \surnamestart Bouzafour\surnameend},
  \bibinfo{author}{Marc \surnamestart Renaudin\surnameend},
  \bibinfo{author}{Hubert \surnamestart Garavel\surnameend},
  \bibinfo{author}{Radu \surnamestart Mateescu\surnameend} \&
  \bibinfo{author}{Wendelin \surnamestart Serwe\surnameend}
  (\bibinfo{year}{2018}): \emph{\bibinfo{title}{{Model-checking Synthesizable
  SystemVerilog Descriptions of Asynchronous Circuits}}}.
\newblock In \bibinfo{editor}{Milos \surnamestart Krstic\surnameend} \&
  \bibinfo{editor}{Ian~W. \surnamestart Jones\surnameend}, editors: {\sl
  \bibinfo{booktitle}{Proceedings of the 24th IEEE International Symposium on
  Asynchronous Circuits and Systems (ASYNC'18), Vienna, Austria}},
  \bibinfo{publisher}{IEEE}, pp. \bibinfo{pages}{34--42},
  \doi{10.1109/ASYNC.2018.00021}.

\bibitemdeclare{article}{Brookes-Hoare-Roscoe-84}
\bibitem{Brookes-Hoare-Roscoe-84}
\bibinfo{author}{Stephen~D. \surnamestart Brookes\surnameend},
  \bibinfo{author}{C.~A.~R. \surnamestart Hoare\surnameend} \&
  \bibinfo{author}{A.~W. \surnamestart Roscoe\surnameend}
  (\bibinfo{year}{1984}): \emph{\bibinfo{title}{{A Theory of Communicating
  Sequential Processes}}}.
\newblock {\sl \bibinfo{journal}{Journal of the ACM}}
  \bibinfo{volume}{31}(\bibinfo{number}{3}), pp. \bibinfo{pages}{560--599},
  \doi{10.1145/828.833}.

\bibitemdeclare{unpublished}{Champelovier-Clerc-Garavel-et-al-10-v6.8}
\bibitem{Champelovier-Clerc-Garavel-et-al-10-v6.8}
\bibinfo{author}{David \surnamestart Champelovier\surnameend},
  \bibinfo{author}{Xavier \surnamestart Clerc\surnameend},
  \bibinfo{author}{Hubert \surnamestart Garavel\surnameend},
  \bibinfo{author}{Yves \surnamestart Guerte\surnameend},
  \bibinfo{author}{Christine \surnamestart McKinty\surnameend},
  \bibinfo{author}{Vincent \surnamestart Powazny\surnameend},
  \bibinfo{author}{Fr\'ed\'eric \surnamestart Lang\surnameend},
  \bibinfo{author}{Wendelin \surnamestart Serwe\surnameend} \&
  \bibinfo{author}{Gideon \surnamestart Smeding\surnameend}
  (\bibinfo{year}{2019}): \emph{\bibinfo{title}{{Reference Manual of the LNT to
  LOTOS Translator (Version 6.8)}}}.
\newblock
  \urlprefix\url{http://cadp.inria.fr/publications/Champelovier-Clerc-Garavel-et-al-10.html}.
\newblock \bibinfo{note}{{INRIA}, Grenoble, France}.

\bibitemdeclare{inproceedings}{Cleaveland-Parrow-Steffen-89}
\bibitem{Cleaveland-Parrow-Steffen-89}
\bibinfo{author}{Rance \surnamestart Cleaveland\surnameend},
  \bibinfo{author}{Joachim \surnamestart Parrow\surnameend} \&
  \bibinfo{author}{Bernhard \surnamestart Steffen\surnameend}
  (\bibinfo{year}{1989}): \emph{\bibinfo{title}{{The Concurrency Workbench}}}.
\newblock In \bibinfo{editor}{Joseph \surnamestart Sifakis\surnameend}, editor:
  {\sl \bibinfo{booktitle}{Proceedings of the 1st Workshop on Automatic
  Verification Methods for Finite State Systems, Grenoble, France}}, {\sl
  \bibinfo{series}{Lecture Notes in Computer Science}} \bibinfo{volume}{407},
  \bibinfo{publisher}{Springer Verlag}, pp. \bibinfo{pages}{24--37},
  \doi{10.1007/3-540-52148-8_3}.

\bibitemdeclare{phdthesis}{Dill-88}
\bibitem{Dill-88}
\bibinfo{author}{David~L. \surnamestart Dill\surnameend}
  (\bibinfo{year}{1988}): \emph{\bibinfo{title}{{Trace Theory for Automatic
  Hierarchical Verification of Speed-Independent Circuits}}}.
\newblock \bibinfo{type}{Acm distinguished dissertation},
  \bibinfo{school}{Carnegie Mellon University, Pittsburgh, PA, USA}.
\newblock
  \urlprefix\url{http://reports-archive.adm.cs.cmu.edu/anon/scan/CMU-CS-88-119.pdf}.

\bibitemdeclare{article}{Edwards-Bardsley-02}
\bibitem{Edwards-Bardsley-02}
\bibinfo{author}{Doug \surnamestart Edwards\surnameend} \&
  \bibinfo{author}{Andrew \surnamestart Bardsley\surnameend}
  (\bibinfo{year}{2002}): \emph{\bibinfo{title}{{Balsa: An Asynchronous
  Hardware Synthesis Language}}}.
\newblock {\sl \bibinfo{journal}{The Computer Journal}}
  \bibinfo{volume}{45}(\bibinfo{number}{1}), pp. \bibinfo{pages}{12--18},
  \doi{10.1093/comjnl/45.1.12}.

\bibitemdeclare{inproceedings}{Garavel-Lang-01}
\bibitem{Garavel-Lang-01}
\bibinfo{author}{Hubert \surnamestart Garavel\surnameend} \&
  \bibinfo{author}{Fr\'ed\'eric \surnamestart Lang\surnameend}
  (\bibinfo{year}{2001}): \emph{\bibinfo{title}{{SVL: a Scripting Language for
  Compositional Verification}}}.
\newblock In \bibinfo{editor}{Myungchul \surnamestart Kim\surnameend},
  \bibinfo{editor}{Byoungmoon \surnamestart Chin\surnameend},
  \bibinfo{editor}{Sungwon \surnamestart Kang\surnameend} \&
  \bibinfo{editor}{Danhyung \surnamestart Lee\surnameend}, editors: {\sl
  \bibinfo{booktitle}{Proceedings of the 21st IFIP WG 6.1 International
  Conference on Formal Techniques for Networked and Distributed Systems
  (FORTE'01), Cheju Island, Korea}}, \bibinfo{publisher}{Kluwer Academic
  Publishers}, pp. \bibinfo{pages}{377--392}.
\newblock
  \urlprefix\url{http://cadp.inria.fr/publications/Garavel-Lang-01.html}.

\bibitemdeclare{article}{Garavel-Lang-Mateescu-Serwe-13}
\bibitem{Garavel-Lang-Mateescu-Serwe-13}
\bibinfo{author}{Hubert \surnamestart Garavel\surnameend},
  \bibinfo{author}{Fr\'ed\'eric \surnamestart Lang\surnameend},
  \bibinfo{author}{Radu \surnamestart Mateescu\surnameend} \&
  \bibinfo{author}{Wendelin \surnamestart Serwe\surnameend}
  (\bibinfo{year}{2013}): \emph{\bibinfo{title}{{CADP 2011: A Toolbox for the
  Construction and Analysis of Distributed Processes}}}.
\newblock {\sl \bibinfo{journal}{Springer International Journal on Software
  Tools for Technology Transfer (STTT)}}
  \bibinfo{volume}{15}(\bibinfo{number}{2}), pp. \bibinfo{pages}{89--107},
  \doi{10.1007/s10009-012-0244-z}.

\bibitemdeclare{inproceedings}{Garavel-Lang-Serwe-17}
\bibitem{Garavel-Lang-Serwe-17}
\bibinfo{author}{Hubert \surnamestart Garavel\surnameend},
  \bibinfo{author}{Fr\'ed\'eric \surnamestart Lang\surnameend} \&
  \bibinfo{author}{Wendelin \surnamestart Serwe\surnameend}
  (\bibinfo{year}{2017}): \emph{\bibinfo{title}{{From LOTOS to LNT}}}.
\newblock In \bibinfo{editor}{Joost-Pieter \surnamestart Katoen\surnameend},
  \bibinfo{editor}{Rom \surnamestart Langerak\surnameend} \&
  \bibinfo{editor}{Arend \surnamestart Rensink\surnameend}, editors: {\sl
  \bibinfo{booktitle}{ModelEd, TestEd, TrustEd -- Essays Dedicated to Ed
  Brinksma on the Occasion of His 60th Birthday}}, {\sl
  \bibinfo{series}{Lecture Notes in Computer Science}} \bibinfo{volume}{10500},
  \bibinfo{publisher}{Springer Verlag}, pp. \bibinfo{pages}{3--26},
  \doi{10.1007/978-3-319-68270-9_1}.

\bibitemdeclare{inproceedings}{Garavel-Serwe-17}
\bibitem{Garavel-Serwe-17}
\bibinfo{author}{Hubert \surnamestart Garavel\surnameend} \&
  \bibinfo{author}{Wendelin \surnamestart Serwe\surnameend}
  (\bibinfo{year}{2017}): \emph{\bibinfo{title}{{The Unheralded Value of the
  Multiway Rendezvous: Illustration with the Production Cell Benchmark}}}.
\newblock In \bibinfo{editor}{Holger \surnamestart Hermanns\surnameend} \&
  \bibinfo{editor}{Peter \surnamestart H\"{o}fner\surnameend}, editors: {\sl
  \bibinfo{booktitle}{Proceedings of the 2nd Workshop on Models for Formal
  Analysis of Real Systems (MARS'17), Uppsala, Sweden}}, {\sl
  \bibinfo{series}{Electronic Proceedings in Computer Science}}
  \bibinfo{volume}{244}, pp. \bibinfo{pages}{230--270},
  \doi{10.4204/EPTCS.244.10}.

\bibitemdeclare{book}{Hoare-85}
\bibitem{Hoare-85}
\bibinfo{author}{C.~A.~R. \surnamestart Hoare\surnameend}
  (\bibinfo{year}{1985}): \emph{\bibinfo{title}{{Communicating Sequential
  Processes}}}.
\newblock \bibinfo{publisher}{Prentice-Hall}.
\newblock \urlprefix\url{http://usingcsp.com/cspbook.pdf}.

\bibitemdeclare{inproceedings}{Jia-Li-He-17}
\bibitem{Jia-Li-He-17}
\bibinfo{author}{Tingting \surnamestart Jia\surnameend},
  \bibinfo{author}{Caihong \surnamestart Li\surnameend} \&
  \bibinfo{author}{Anping \surnamestart He\surnameend} (\bibinfo{year}{2017}):
  \emph{\bibinfo{title}{{Modeling and Verification of Circuit with
  Stable-Event}}}.
\newblock In: {\sl \bibinfo{booktitle}{Proceedings of the 2017 International
  Conference on Cyber-Enabled Distributed Computing and Knowledge Discovery
  (CyberC), Nanjing, China}}, \bibinfo{publisher}{IEEE}, pp.
  \bibinfo{pages}{471--475}, \doi{10.1109/CyberC.2017.73}.

\bibitemdeclare{article}{Josephs-92}
\bibitem{Josephs-92}
\bibinfo{author}{Mark~B. \surnamestart Josephs\surnameend}
  (\bibinfo{year}{1992}): \emph{\bibinfo{title}{{Receptive Process Theory}}}.
\newblock {\sl \bibinfo{journal}{Acta Informatica}}
  \bibinfo{volume}{29}(\bibinfo{number}{1}), pp. \bibinfo{pages}{17--31},
  \doi{10.1007/BF01178564}.

\bibitemdeclare{inproceedings}{Josephs-07}
\bibitem{Josephs-07}
\bibinfo{author}{Mark~B. \surnamestart Josephs\surnameend}
  (\bibinfo{year}{2007}): \emph{\bibinfo{title}{{Gate-level Modelling and
  Verification of Asynchronous Circuits using CSPM and FDR}}}.
\newblock In: {\sl \bibinfo{booktitle}{Proceedings of the 13th IEEE
  International Symposium on Asynchronous Circuits and Systems (ASYNC'07),
  Berkeley, California, USA}}, \bibinfo{publisher}{IEEE}, pp.
  \bibinfo{pages}{83--94}, \doi{10.1109/ASYNC.2007.19}.

\bibitemdeclare{article}{Kapoor-Josephs-04}
\bibitem{Kapoor-Josephs-04}
\bibinfo{author}{Hemangee~K. \surnamestart Kapoor\surnameend} \&
  \bibinfo{author}{Mark~B. \surnamestart Josephs\surnameend}
  (\bibinfo{year}{2004}): \emph{\bibinfo{title}{{Modelling and Verification of
  Delay-Insensitive Circuits using CCS and the Concurrency Workbench}}}.
\newblock {\sl \bibinfo{journal}{Information Processing Letters}}
  \bibinfo{volume}{89}(\bibinfo{number}{6}), pp. \bibinfo{pages}{293--296},
  \doi{10.1016/j.ipl.2003.12.007}.

\bibitemdeclare{inproceedings}{Lang-05}
\bibitem{Lang-05}
\bibinfo{author}{Fr\'ed\'eric \surnamestart Lang\surnameend}
  (\bibinfo{year}{2005}): \emph{\bibinfo{title}{{EXP.OPEN 2.0: A Flexible Tool
  Integrating Partial Order, Compositional, and On-the-fly Verification
  Methods}}}.
\newblock In \bibinfo{editor}{Judi \surnamestart Romijn\surnameend},
  \bibinfo{editor}{Graeme \surnamestart Smith\surnameend} \&
  \bibinfo{editor}{Jaco \surnamestart {van de Pol}\surnameend}, editors: {\sl
  \bibinfo{booktitle}{Proceedings of the 5th International Conference on
  Integrated Formal Methods (IFM'05), Eindhoven, The Netherlands}}, {\sl
  \bibinfo{series}{Lecture Notes in Computer Science}} \bibinfo{volume}{3771},
  \bibinfo{publisher}{Springer Verlag}, pp. \bibinfo{pages}{70--88},
  \doi{10.1007/11589976\_6}.

\bibitemdeclare{article}{Martin-86}
\bibitem{Martin-86}
\bibinfo{author}{Alain~J. \surnamestart Martin\surnameend}
  (\bibinfo{year}{1986}): \emph{\bibinfo{title}{{Compiling Communicating
  Processes into Delay-Insensitive VLSI Circuits}}}.
\newblock {\sl \bibinfo{journal}{Distributed Computing}}
  \bibinfo{volume}{1}(\bibinfo{number}{4}), pp. \bibinfo{pages}{226--234},
  \doi{10.1007/BF01660034}.

\bibitemdeclare{techreport}{Martin-14}
\bibitem{Martin-14}
\bibinfo{author}{Alain~J. \surnamestart Martin\surnameend}
  (\bibinfo{year}{2014}): \emph{\bibinfo{title}{{25 Years Ago: The First
  Asynchronous Microprocessor}}}.
\newblock \bibinfo{type}{Computer Science Technical Reports}
  \bibinfo{number}{2014.001}, \bibinfo{institution}{California Institute of
  Technology, Pasadena, California, USA}.
\newblock
  \urlprefix\url{https://resolver.caltech.edu/CaltechAUTHORS:20140206-111915844}.

\bibitemdeclare{techreport}{Muller-55}
\bibitem{Muller-55}
\bibinfo{author}{David~E. \surnamestart Muller\surnameend}
  (\bibinfo{year}{1955}): \emph{\bibinfo{title}{{Theory of Asynchronous
  Circuits}}}.
\newblock \bibinfo{type}{Research report} \bibinfo{number}{66},
  \bibinfo{institution}{University of Illinois at Urbana-Champaign, Department
  of Computer Science}.
\newblock \urlprefix\url{https://archive.org/details/theoryofasynchro66mull}.

\bibitemdeclare{article}{Plana-Riocreux-Bainbridge-et-al-03}
\bibitem{Plana-Riocreux-Bainbridge-et-al-03}
\bibinfo{author}{Luis~A. \surnamestart Plana\surnameend},
  \bibinfo{author}{P.~A. \surnamestart Riocreux\surnameend},
  \bibinfo{author}{W.~J. \surnamestart Bainbridge\surnameend},
  \bibinfo{author}{Andrew \surnamestart Bardsley\surnameend},
  \bibinfo{author}{Steve \surnamestart Temple\surnameend},
  \bibinfo{author}{Jim~D. \surnamestart Garside\surnameend} \&
  \bibinfo{author}{Z.~C. \surnamestart Yu\surnameend} (\bibinfo{year}{2003}):
  \emph{\bibinfo{title}{{SPA -- a secure Amulet core for smartcard
  applications}}}.
\newblock {\sl \bibinfo{journal}{Microprocessors and Microsystems}}
  \bibinfo{volume}{27}(\bibinfo{number}{9}), pp. \bibinfo{pages}{431--446},
  \doi{10.1016/S0141-9331(03)00093-0}.

\bibitemdeclare{techreport}{INPI-FR-3054344}
\bibitem{INPI-FR-3054344}
\bibinfo{author}{Marc \surnamestart Renaudin\surnameend},
  \bibinfo{author}{Bertrand \surnamestart Folco\surnameend} \&
  \bibinfo{author}{Boulahia \surnamestart Boubkar\surnameend}
  (\bibinfo{year}{2018}): \emph{\bibinfo{title}{{Circuit int\'egr\'e
  prot\'eg\'e}}}.
\newblock \bibinfo{type}{Brevet d'invention 16~57129}
  \bibinfo{number}{3~054~344}, \bibinfo{institution}{Institut National de la
  Propri\'et\'e Industrielle}.

\bibitemdeclare{techreport}{EPO-EP-3276656-B1}
\bibitem{EPO-EP-3276656-B1}
\bibinfo{author}{Marc \surnamestart Renaudin\surnameend},
  \bibinfo{author}{Bertrand \surnamestart Folco\surnameend} \&
  \bibinfo{author}{Boulahia \surnamestart Boubkar\surnameend}
  (\bibinfo{year}{2019}): \emph{\bibinfo{title}{{Circuit int\'egr\'e
  prot\'eg\'e}}}.
\newblock \bibinfo{type}{Fascicule de Brevet Europeen}
  \bibinfo{number}{EP~3~276~656~B1}, \bibinfo{institution}{European Patent
  Office}.

\bibitemdeclare{book}{Roscoe-Hoare-Bird-97}
\bibitem{Roscoe-Hoare-Bird-97}
\bibinfo{author}{A.~W. \surnamestart Roscoe\surnameend}, \bibinfo{author}{{C.
  A. R.} \surnamestart Hoare\surnameend} \& \bibinfo{author}{Richard
  \surnamestart Bird\surnameend} (\bibinfo{year}{1997}):
  \emph{\bibinfo{title}{{The Theory and Practice of Concurrency}}}.
\newblock \bibinfo{publisher}{Prentice Hall}.
\newblock \urlprefix\url{https://archive.org/details/theorypracticeof00rosc}.

\bibitemdeclare{inproceedings}{Stevens-Aldwinckle-Birtwistle-Liu-93}
\bibitem{Stevens-Aldwinckle-Birtwistle-Liu-93}
\bibinfo{author}{Ken \surnamestart Stevens\surnameend}, \bibinfo{author}{John
  \surnamestart Aldwinckle\surnameend}, \bibinfo{author}{Graham \surnamestart
  Birtwistle\surnameend} \& \bibinfo{author}{Ying \surnamestart Liu\surnameend}
  (\bibinfo{year}{1993}): \emph{\bibinfo{title}{Designing Parallel
  Specifications in {CCS}}}.
\newblock In: {\sl \bibinfo{booktitle}{Proceedings of the Canadian Conference
  on Electrical and Computer Engineering}}, pp. \bibinfo{pages}{983--986},
  \doi{10.1109/CCECE.1993.332460}.

\bibitemdeclare{inproceedings}{Verhoeff-98}
\bibitem{Verhoeff-98}
\bibinfo{author}{Tom \surnamestart Verhoeff\surnameend} (\bibinfo{year}{1998}):
  \emph{\bibinfo{title}{{Analyzing Specifications for Delay-Insensitive
  Circuits}}}.
\newblock In: {\sl \bibinfo{booktitle}{Proceedings of the Fourth International
  Symposium on Advanced Research in Asynchronous Circuits and Systems
  (ASYNC'98), San Diego, California, USA}}, \bibinfo{publisher}{IEEE}, pp.
  \bibinfo{pages}{172--183}, \doi{10.1109/ASYNC.1998.666503}.

\bibitemdeclare{inproceedings}{Wang-Kwiatkowska-et-al-06}
\bibitem{Wang-Kwiatkowska-et-al-06}
\bibinfo{author}{X.~\surnamestart Wang\surnameend},
  \bibinfo{author}{M.~\surnamestart Kwiatkowska\surnameend},
  \bibinfo{author}{G.~\surnamestart Theodoropoulos\surnameend} \&
  \bibinfo{author}{Q.~\surnamestart Zhang\surnameend} (\bibinfo{year}{2006}):
  \emph{\bibinfo{title}{Opportunities and Challenges in Process-Algebraic
  Verification of Asynchronous Circuit Designs}}.
\newblock In: {\sl \bibinfo{booktitle}{Proceedings of the Second Workshop on
  Globally Asynchronous Locally Synchronous Design (FMGALS'05)}}, {\sl
  \bibinfo{series}{Electronic Notes in Theoretical Computer Science}}
  \bibinfo{volume}{146}, pp. \bibinfo{pages}{189--206},
  \doi{10.1016/j.entcs.2005.05.042}.

\bibitemdeclare{article}{Wang-Kwiatkowska-07}
\bibitem{Wang-Kwiatkowska-07}
\bibinfo{author}{Xu~\surnamestart Wang\surnameend} \& \bibinfo{author}{Marta~Z.
  \surnamestart Kwiatkowska\surnameend} (\bibinfo{year}{2007}):
  \emph{\bibinfo{title}{{On Process-algebraic Verification of Asynchronous
  Circuits}}}.
\newblock {\sl \bibinfo{journal}{Fundamenta Informaticae}}
  \bibinfo{volume}{80}(\bibinfo{number}{1--3}), pp. \bibinfo{pages}{283--310}.
\newblock
  \urlprefix\url{https://content.iospress.com/download/fundamenta-informaticae/fi80-1-3-16?id=fundamenta-informaticae%2Ffi80-1-3-16}.

\bibitemdeclare{article}{Yakovlev-Kishinevsky-Kondratyev-et-al-96}
\bibitem{Yakovlev-Kishinevsky-Kondratyev-et-al-96}
\bibinfo{author}{Alexandre \surnamestart Yakovlev\surnameend},
  \bibinfo{author}{Michael \surnamestart Kishinevsky\surnameend},
  \bibinfo{author}{Alex \surnamestart Kondratyev\surnameend},
  \bibinfo{author}{Luciano \surnamestart Lavagno\surnameend} \&
  \bibinfo{author}{Marta \surnamestart Pietkiewicz-Koutny\surnameend}
  (\bibinfo{year}{1996}): \emph{\bibinfo{title}{{On the Models for Asynchronous
  Circuit Behaviour with OR Causality}}}.
\newblock {\sl \bibinfo{journal}{Formal Methods in System Design}}
  \bibinfo{volume}{9}(\bibinfo{number}{3}), pp. \bibinfo{pages}{189--233},
  \doi{10.1007/BF00122082}.

\bibitemdeclare{article}{Yakovlev-Koelmans-Semenov-et-al-96}
\bibitem{Yakovlev-Koelmans-Semenov-et-al-96}
\bibinfo{author}{Alexandre~V. \surnamestart Yakovlev\surnameend},
  \bibinfo{author}{Albert~M. \surnamestart Koelmans\surnameend},
  \bibinfo{author}{Alexei~L. \surnamestart Semenov\surnameend} \&
  \bibinfo{author}{David~J. \surnamestart Kinniment\surnameend}
  (\bibinfo{year}{1996}): \emph{\bibinfo{title}{{Modelling, Analysis and
  Synthesis of Asynchronous Control Circuits using Petri Nets}}}.
\newblock {\sl \bibinfo{journal}{Integration}}
  \bibinfo{volume}{21}(\bibinfo{number}{3}), pp. \bibinfo{pages}{143--170},
  \doi{10.1016/S0167-9260(96)00010-7}.

\end{thebibliography}

\appendix
\section{LNT Models}
\label{app:lnt}

The LNT model of the shield takes advantage of modules~\cite{Champelovier-Clerc-Garavel-et-al-10-v6.8} to split the overall model into different files: a module with the definition of basic data types, a (family of) module(s) for the gates, a module defining stub processes, and the main module.

\subsection{Data Types, Operations, and Channels}
\label{app:lnt:types}

Module \lstinline+VOLTAGE+ defines an enumerated data type for voltages, logical operations (used in the gates), channels for communicating voltages, and processes to represent wires and (isochronic) forks.

\begin{lstlisting}[firstline=5]
module VOLTAGE is

type VOLTAGE is
   DOWN, UP
   with "==", "!="
end type

------------------------------------------------------------------------

function NOT (X: VOLTAGE) : VOLTAGE is
   if X == DOWN then
      return UP
   else
      return DOWN
   end if
end function

function _AND_ (X1, X2: VOLTAGE) : VOLTAGE is
   case X1, X2 in
      UP, UP   -> return UP
    | any, any -> return DOWN
   end case
end function

function _OR_ (X1, X2: VOLTAGE) : VOLTAGE is
   case X1, X2 in
      DOWN, DOWN -> return DOWN
    | any, any   -> return UP
   end case
end function

function MULLER (X1, X2, PREVIOUS: VOLTAGE) : VOLTAGE is
   if (X1 == DOWN) and (X2 == DOWN) then
      return DOWN
   elsif (X1 == UP) and (X2 == UP) then
      return UP
   else
      return PREVIOUS
   end if
end function

------------------------------------------------------------------------

channel LINK is
   (VOLTAGE)
end channel

------------------------------------------------------------------------
-- straight wire (with non-zero delay)
-- also used for modeling an isochronic fork

process WIRE [INPUT, OUTPUT: LINK] is
   var X : VOLTAGE in
      loop
         INPUT (?X);
      	 OUTPUT (X)
      end loop
   end var
end process

------------------------------------------------------------------------
-- forking wire replicating an input on both outputs in any order

process FORK [INPUT, OUTPUT1, OUTPUT2: LINK] is
   var X : VOLTAGE in
      loop
         INPUT (?X);
         par
            OUTPUT1 (X)
         || OUTPUT2 (X)
         end par
      end loop
   end var
end process

------------------------------------------------------------------------
-- wire stuck at a given voltage

process STUCKAT [W: LINK] (V: VOLTAGE) is
   loop
      W (V)
   end loop
end process

end module
\end{lstlisting}

\subsection{Gates}
\label{app:lnt:gates}

For each modeling style (see Fig.~\ref{fig:and}), there is dedicated version of the module \lstinline+GATES+.

\subsubsection{Transition-Oriented Gates}
\label{app:lnt:gates:transition}

\begin{lstlisting}[firstline=4]
module GATES (VOLTAGE) is

process INV [INPUT, OUTPUT: LINK] (in var X: VOLTAGE) is
   -- inverter
   var RESULT, NEW_RESULT: VOLTAGE in
      RESULT := NOT (X);
      loop
         INPUT (?X);
         NEW_RESULT := NOT (X);
         if NEW_RESULT != RESULT then
            RESULT := NEW_RESULT;
            OUTPUT (RESULT)
         end if
      end loop
   end var
end process

process AND [INPUT1, INPUT2, OUTPUT: LINK] (in var X1, X2: VOLTAGE) is
   -- binary AND gate
   var RESULT, NEW_RESULT: VOLTAGE in
      RESULT := X1 AND X2;
      loop
         select
            INPUT1 (?X1)
         [] INPUT2 (?X2)
         end select;
         NEW_RESULT := X1 AND X2; 
         if NEW_RESULT != RESULT then
            RESULT := NEW_RESULT;
            OUTPUT (RESULT)
         end if
      end loop
   end var
end process

process NOR [INPUT1, INPUT2, OUTPUT: LINK] (in var X1, X2: VOLTAGE) is
   -- binary NOR gate
   var RESULT, NEW_RESULT: VOLTAGE in
      RESULT := NOT (X1 OR X2);
      loop
         select
            INPUT1 (?X1)
         [] INPUT2 (?X2)
         end select;
         NEW_RESULT := NOT (X1 OR X2);
         if NEW_RESULT != RESULT then
            RESULT := NEW_RESULT;
            OUTPUT (RESULT)
         end if
      end loop
   end var
end process

process MULLER [INPUT1, INPUT2, OUTPUT: LINK]
               (in var X1, X2, RESULT: VOLTAGE) is
   -- Muller's C element
   var NEW_RESULT: VOLTAGE in
      loop
         select
            INPUT1 (?X1)
         [] INPUT2 (?X2)
         end select;
         NEW_RESULT := MULLER (X1, X2, RESULT);
         if NEW_RESULT != RESULT then
            RESULT := NEW_RESULT;
            OUTPUT (RESULT)
         end if
      end loop
   end var      
end process

end module
\end{lstlisting}

\subsection{Intuitive State-Oriented Gates}
\label{app:lnt:gates:intuitive}

\begin{lstlisting}[firstline=5]
module GATES (VOLTAGE) is

process INV [INPUT, OUTPUT: LINK] (in var X: VOLTAGE) is
   -- inverter
   loop
      INPUT (?X);
      OUTPUT (NOT (X))
   end loop
end process

process BINARY [INPUT1, INPUT2: LINK] (in out X1, X2: VOLTAGE) is
   -- accept one of two inputs
   use X1, X2; -- to keep lnt2lotos silent
   select
      INPUT1 (?X1)
   [] INPUT2 (?X2)
   end select
end process

process AND [INPUT1, INPUT2, OUTPUT: LINK] (in var X1, X2: VOLTAGE) is
   -- binary AND gate
   loop
      BINARY [INPUT1, INPUT2] (!?X1, !?X2);
      OUTPUT (X1 AND X2)
   end loop
end process

process NOR [INPUT1, INPUT2, OUTPUT: LINK] (in var X1, X2: VOLTAGE) is
   -- binary NOR gate
   loop
      BINARY [INPUT1, INPUT2] (!?X1, !?X2);
      OUTPUT (NOT (X1 OR X2))
   end loop
end process

process MULLER [INPUT1, INPUT2, OUTPUT: LINK]
               (in var X1, X2, RESULT: VOLTAGE) is
   -- Muller's C element
   loop
      BINARY [INPUT1, INPUT2] (!?X1, !?X2);
      RESULT := MULLER (X1, X2, RESULT);
      OUTPUT (RESULT)
   end loop
end process

end module
\end{lstlisting}

\subsubsection{State-Oriented Gates}
\label{app:lnt:gates:state}

\begin{lstlisting}[firstline=4]
module GATES (VOLTAGE) is

process INV [INPUT, OUTPUT: LINK] (in var X: VOLTAGE) is
   -- inverter
   loop
      INPUT (?X);
      OUTPUT (NOT (X))
   end loop
end process

process BINARY [INPUT1, INPUT2: LINK] (in out X1, X2: VOLTAGE) is
   -- accept one or two inputs in arbitrary order
   use X1, X2; -- to keep lnt2lotos silent
   select
      INPUT1 (?X1);
      select
         null
      [] INPUT2 (?X2)
      end select
   [] INPUT2 (?X2);
      select
         null
      [] INPUT1 (?X1)
      end select
   end select
end process

process AND [INPUT1, INPUT2, OUTPUT: LINK] (in var X1, X2: VOLTAGE) is
   -- binary AND gate
   loop
      BINARY [INPUT1, INPUT2] (!?X1, !?X2);
      OUTPUT (X1 AND X2)
   end loop
end process

process NOR [INPUT1, INPUT2, OUTPUT: LINK] (in var X1, X2: VOLTAGE) is
   -- binary NOR gate
   loop
      BINARY [INPUT1, INPUT2] (!?X1, !?X2);
      OUTPUT (NOT (X1 OR X2))
   end loop
end process

process MULLER [INPUT1, INPUT2, OUTPUT: LINK]
               (in var X1, X2, RESULT: VOLTAGE) is
   -- Muller's C element
   loop
      BINARY [INPUT1, INPUT2] (!?X1, !?X2);
      RESULT := MULLER (X1, X2, RESULT);
      OUTPUT (RESULT)
   end loop
end process

end module
\end{lstlisting}

\subsubsection{Parallel State-Oriented Gates}
\label{app:lnt:gates:parallel}

\begin{lstlisting}[firstline=5]
module GATES (VOLTAGE) is

process INV [INPUT, OUTPUT: LINK] (in var X: VOLTAGE) is
   -- inverter
   loop
      select
         INPUT (?X)
      [] null
      end select;
      OUTPUT (NOT (X))
   end loop
end process

process BINARY [INPUT1, INPUT2: LINK] (in out X1, X2: VOLTAGE) is
   -- accept one or two inputs in arbitrary order
   use X1, X2; -- to keep lnt2lotos silent
   par
      select
         INPUT1 (?X1)
      [] null
      end select 
   || select
         INPUT2 (?X2)
      [] null
      end select
   end par
end process

process AND [INPUT1, INPUT2, OUTPUT: LINK] (in var X1, X2: VOLTAGE) is
   -- binary AND gate
   loop
      BINARY [INPUT1, INPUT2] (!?X1, !?X2);
      OUTPUT (X1 AND X2)
   end loop
end process

process NOR [INPUT1, INPUT2, OUTPUT: LINK] (in var X1, X2: VOLTAGE) is
   -- binary NOR gate
   loop
      BINARY [INPUT1, INPUT2] (!?X1, !?X2);
      OUTPUT (NOT (X1 OR X2))
   end loop
end process

process MULLER [INPUT1, INPUT2, OUTPUT: LINK]
               (in var X1, X2, RESULT: VOLTAGE) is
   -- Muller's C element
   loop
      BINARY [INPUT1, INPUT2] (!?X1, !?X2);
      RESULT := MULLER (X1, X2, RESULT);
      OUTPUT (RESULT)
   end loop
end process

end module
\end{lstlisting}

\subsubsection{Free State-Oriented Gates}
\label{app:lnt:gates:free}

\begin{lstlisting}[firstline=6]
module GATES (VOLTAGE) is

process INV [INPUT, OUTPUT: LINK] (in var X: VOLTAGE) is
   -- inverter
   loop
      select
         INPUT (?X)
      [] OUTPUT (NOT (X))
      end select
   end loop
end process

process AND [INPUT1, INPUT2, OUTPUT: LINK] (in var X1, X2: VOLTAGE) is
   -- binary AND gate
   loop
      select
         INPUT1 (?X1)
      [] INPUT2 (?X2)
      [] OUTPUT (X1 AND X2)
      end select
   end loop
end process

process NOR [INPUT1, INPUT2, OUTPUT: LINK] (in var X1, X2: VOLTAGE) is
   -- binary NOR gate
   loop
      select
         INPUT1 (?X1)
      [] INPUT2 (?X2)
      [] OUTPUT (NOT (X1 OR X2))
      end select
   end loop
end process

process MULLER [INPUT1, INPUT2, OUTPUT: LINK]
               (in var X1, X2, RESULT: VOLTAGE) is
   -- Muller's C element
   loop
      select
         INPUT1 (?X1)
      [] INPUT2 (?X2)
      [] OUTPUT (RESULT)
      end select;
      RESULT := MULLER (X1, X2, RESULT)
   end loop
end process

end module
\end{lstlisting}

\subsection{Stubs}
\label{app:lnt:stubs}

\begin{lstlisting}[firstline=5]
module STUBS (VOLTAGE) is

-----------------------------------------------------------------------
-- behaviour (protocol) of an asynchronous sequencer as given by a
-- (sequential) signal transition graph

process PROTOCOL [R_PRED, A_PRED, R_SUCC, A_SUCC: LINK] is
   loop
      R_PRED (UP);
      R_SUCC (UP);
      A_SUCC (UP);
      R_SUCC (DOWN);
      A_SUCC (DOWN);
      A_PRED (UP);
      R_PRED (DOWN);
      A_PRED (DOWN)
   end loop
end process

-----------------------------------------------------------------------
-- process to absorb repeated inputs, transforming a state-oriented
-- model into a transition-oriented one

process ABSORB [W: LINK] (V: VOLTAGE) is
   loop L in
      select
         W (V)
      [] break L
      end select
   end loop
end process

-----------------------------------------------------------------------
-- stub processes enforcing the state-oriented protocol with the same
-- gates as the sequencer so as to use the same composition operator
-- to add a stub instead of adding a sequencer
-- contrary to the transition-oriented protocol, repeated outputs are
-- permitted: thus, left and right stubs are different

process STUB_L [R_PRED, A_PRED, R_SUCC, A_SUCC: LINK] is
   var V: VOLTAGE in
      V := UP;
      loop
	 ABSORB [A_SUCC] (NOT (V));
         R_PRED (V); R_SUCC (V); -- immediate propagation
         ABSORB [A_SUCC] (NOT (V));
         A_SUCC (V);
         A_PRED (V);
         V := NOT (V)
      end loop
   end var
end process

process STUB_R [R_PRED, A_PRED, R_SUCC, A_SUCC: LINK] is
   var V: VOLTAGE in
      V := UP;
      loop
         ABSORB [R_PRED] (NOT (V));
         R_PRED (V);
         R_SUCC (V);
	 ABSORB [R_PRED] (V);
         A_SUCC (V); A_PRED (V); -- immediate propagation
         V := NOT (V)
      end loop
   end var
end process

end module
\end{lstlisting}

\subsection{Sequencer}
\label{app:lnt:sequencer}

\begin{lstlisting}[firstline=5]
module SEQUENCER (GATES, STUBS) is

------------------------------------------------------------------------
-- behaviour of a sequencer as given by the composition of its
-- elementary gates (AND, NOR, INV, and MULLER), explicitly modeling
-- forks by dedicated processes; an isochronic fork is modeled as a
-- wire with a three-party rendezvous on the output
-- there are eight processes, one for each combination of isochronic
-- and parallel (non-isochronic) forks

process SEQUENCER_PPP [R_PRED, A_PRED, R_SUCC, A_SUCC, G, H,
                       R_PRED1, R_PRED2, A_SUCC1, A_SUCC2,
		       G2, H1, H2: LINK]
                      (X1, X2, INIT_C: VOLTAGE) is
   par
      R_PRED2, A_SUCC1, G ->
         MULLER [R_PRED2, A_SUCC1, G] (X1, X2, INIT_C)
   || R_PRED1, H1 -> AND [R_PRED1, H1, R_SUCC] (X1, NOT (INIT_C))
   || G2, H -> INV [G2, H] (INIT_C)
   || A_SUCC2, H2 -> NOR [A_SUCC2, H2, A_PRED] (X2, NOT (INIT_C))
   || R_PRED1, R_PRED2 -> FORK [R_PRED, R_PRED1, R_PRED2]
   || A_SUCC1, A_SUCC2 -> FORK [A_SUCC, A_SUCC1, A_SUCC2]
   || H, H1, H2 -> FORK [H, H1, H2]
   || G, G2 -> WIRE [G, G2]
   end par
end process

-- - - - - - - - - - - - - - - - - - - - - - - - - - - - - - - - - - - -

process SEQUENCER_PPI [R_PRED, A_PRED, R_SUCC, A_SUCC, G, H,
                       R_PRED1, R_PRED2, A_SUCC1, A_SUCC2,
		       G2, H2: LINK]
                      (X1, X2, INIT_C: VOLTAGE) is
   par
      R_PRED2, A_SUCC1, G ->
         MULLER [R_PRED2, A_SUCC1, G] (X1, X2, INIT_C)
   || R_PRED1, H2 -> AND [R_PRED1, H2, R_SUCC] (X1, NOT (INIT_C))
   || G2, H -> INV [G2, H] (INIT_C)
   || A_SUCC2, H2 -> NOR [A_SUCC2, H2, A_PRED] (X2, NOT (INIT_C))
   || R_PRED1, R_PRED2 -> FORK [R_PRED, R_PRED1, R_PRED2]
   || A_SUCC1, A_SUCC2 -> FORK [A_SUCC, A_SUCC1, A_SUCC2]
   || H, H2 -> WIRE [H, H2]
   || G, G2 -> WIRE [G, G2]
   end par
end process

-- - - - - - - - - - - - - - - - - - - - - - - - - - - - - - - - - - - -

process SEQUENCER_PIP [R_PRED, A_PRED, R_SUCC, A_SUCC, G, H,
                       R_PRED1, R_PRED2, A_SUCC2, G2, H1, H2: LINK]
                      (X1, X2, INIT_C: VOLTAGE) is
   par
      R_PRED2, A_SUCC2, G ->
         MULLER [R_PRED2, A_SUCC2, G] (X1, X2, INIT_C)
   || R_PRED1, H1 -> AND [R_PRED1, H1, R_SUCC] (X1, NOT (INIT_C))
   || G2, H -> INV [G2, H] (INIT_C)
   || A_SUCC2, H2 -> NOR [A_SUCC2, H2, A_PRED] (X2, NOT (INIT_C))
   || R_PRED1, R_PRED2 -> FORK [R_PRED, R_PRED1, R_PRED2]
   || A_SUCC2 -> WIRE [A_SUCC, A_SUCC2]
   || H, H1, H2 -> FORK [H, H1, H2]
   || G, G2 -> WIRE [G, G2]
   end par
end process

-- - - - - - - - - - - - - - - - - - - - - - - - - - - - - - - - - - - -

process SEQUENCER_PII [R_PRED, A_PRED, R_SUCC, A_SUCC, G, H,
                       R_PRED1, R_PRED2, A_SUCC2, G2, H2: LINK]
                      (X1, X2, INIT_C: VOLTAGE) is
   par
      R_PRED2, A_SUCC2, G ->
         MULLER [R_PRED2, A_SUCC2, G] (X1, X2, INIT_C)
   || R_PRED1, H2 -> AND [R_PRED1, H2, R_SUCC] (X1, NOT (INIT_C))
   || G2, H -> INV [G2, H] (INIT_C)
   || A_SUCC2, H2 -> NOR [A_SUCC2, H2, A_PRED] (X2, NOT (INIT_C))
   || R_PRED1, R_PRED2 -> FORK [R_PRED, R_PRED1, R_PRED2]
   || A_SUCC2 -> WIRE [A_SUCC, A_SUCC2]
   || H, H2 -> WIRE [H, H2]
   || G, G2 -> WIRE [G, G2]
   end par
end process

-- - - - - - - - - - - - - - - - - - - - - - - - - - - - - - - - - - - -

process SEQUENCER_IPP [R_PRED, A_PRED, R_SUCC, A_SUCC, G, H,
                       R_PRED2, A_SUCC1, A_SUCC2, G2, H1, H2: LINK]
                      (X1, X2, INIT_C: VOLTAGE) is
   par
      R_PRED2, A_SUCC1, G ->
         MULLER [R_PRED2, A_SUCC1, G] (X1, X2, INIT_C)
   || R_PRED2, H1 -> AND [R_PRED2, H1, R_SUCC] (X1, NOT (INIT_C))
   || G2, H -> INV [G2, H] (INIT_C)
   || A_SUCC2, H2 -> NOR [A_SUCC2, H2, A_PRED] (X2, NOT (INIT_C))
   || R_PRED2 -> WIRE [R_PRED, R_PRED2]
   || A_SUCC1, A_SUCC2 -> FORK [A_SUCC, A_SUCC1, A_SUCC2]
   || H, H1, H2 -> FORK [H, H1, H2]
   || G, G2 -> WIRE [G, G2]
   end par
end process

-- - - - - - - - - - - - - - - - - - - - - - - - - - - - - - - - - - - -

process SEQUENCER_IPI [R_PRED, A_PRED, R_SUCC, A_SUCC, G, H,
                       R_PRED2, A_SUCC1, A_SUCC2, G2, H2: LINK]
                      (X1, X2, INIT_C: VOLTAGE) is
   par
      R_PRED2, A_SUCC1, G ->
         MULLER [R_PRED2, A_SUCC1, G] (X1, X2, INIT_C)
   || R_PRED2, H2 -> AND [R_PRED2, H2, R_SUCC] (X1, NOT (INIT_C))
   || G2, H -> INV [G2, H] (INIT_C)
   || A_SUCC2, H2 -> NOR [A_SUCC2, H2, A_PRED] (X2, NOT (INIT_C))
   || R_PRED2 -> WIRE [R_PRED, R_PRED2]
   || A_SUCC1, A_SUCC2 -> FORK [A_SUCC, A_SUCC1, A_SUCC2]
   || H, H2 -> WIRE [H, H2]
   || G, G2 -> WIRE [G, G2]
   end par
end process

-- - - - - - - - - - - - - - - - - - - - - - - - - - - - - - - - - - - -

process SEQUENCER_IIP [R_PRED, A_PRED, R_SUCC, A_SUCC, G, H,
                       R_PRED2, A_SUCC2, G2, H1, H2: LINK]
                      (X1, X2, INIT_C: VOLTAGE) is
   par
      R_PRED2, A_SUCC2, G ->
         MULLER [R_PRED2, A_SUCC2, G] (X1, X2, INIT_C)
   || R_PRED2, H1 -> AND [R_PRED2, H1, R_SUCC] (X1, NOT (INIT_C))
   || G2, H -> INV [G2, H] (INIT_C)
   || A_SUCC2, H2 -> NOR [A_SUCC2, H2, A_PRED] (X2, NOT (INIT_C))
   || R_PRED2 -> WIRE [R_PRED, R_PRED2]
   || A_SUCC2 -> WIRE [A_SUCC, A_SUCC2]
   || H, H1, H2 -> FORK [H, H1, H2]
   || G, G2 -> WIRE [G, G2]
   end par
end process

-- - - - - - - - - - - - - - - - - - - - - - - - - - - - - - - - - - - -

process SEQUENCER_III [R_PRED, A_PRED, R_SUCC, A_SUCC, G, H,
                       R_PRED2, A_SUCC2, G2, H2: LINK]
                      (X1, X2, INIT_C: VOLTAGE) is
   par
      R_PRED2, A_SUCC2, G ->
         MULLER [R_PRED2, A_SUCC2, G] (X1, X2, INIT_C)
   || R_PRED2, H2 -> AND [R_PRED2, H2, R_SUCC] (X1, NOT (INIT_C))
   || G2, H -> INV [G2, H] (INIT_C)
   || A_SUCC2, H2 -> NOR [A_SUCC2, H2, A_PRED] (X2, NOT (INIT_C))
   || R_PRED2 -> WIRE [R_PRED, R_PRED2]
   || A_SUCC2 -> WIRE [A_SUCC, A_SUCC2]
   || H, H2 -> WIRE [H, H2]
   || G, G2 -> WIRE [G, G2]
   end par
end process

------------------------------------------------------------------------
-- behaviour of a sequencer as given by the composition of its
-- elementary gates (AND, NOR, INV, and MULLER), using multiway
-- rendezvous to model forking wires, and a gate to model (binary)
-- wires

process SEQUENCER_RV [R_PRED, A_PRED, R_SUCC, A_SUCC, G, H: LINK]
                     (X1, X2, INIT_C: VOLTAGE) is
   par
      R_PRED, A_SUCC, G -> MULLER [R_PRED, A_SUCC, G] (X1, X2, INIT_C)
   || R_PRED, H -> AND [R_PRED, H, R_SUCC] (X1, NOT (INIT_C))
   || G, H -> INV [G, H] (INIT_C)
   || A_SUCC, H -> NOR [A_SUCC, H, A_PRED] (X2, NOT (INIT_C))
   end par
end process

------------------------------------------------------------------------
-- sequencer with the same visible interface as PROTOCOL[]

process SEQUENCER_HIDDEN [R_PRED, A_PRED, R_SUCC, A_SUCC: LINK] is
   hide G, H, R_PRED1, R_PRED2, A_SUCC1, A_SUCC2, G2, H1, H2: LINK in
      SEQUENCER_PPP [R_PRED, A_PRED, R_SUCC, A_SUCC, G, H,
                     R_PRED1, R_PRED2, A_SUCC1, A_SUCC2, G2, H1, H2]
                    (DOWN, DOWN, DOWN)
   end hide
end process

------------------------------------------------------------------------
-- shield with one element (and closed right end)

process SHIELD_1 [R0, A0: LINK] is
   hide R1, A1: LINK in
      par R1, A1 in
         SEQUENCER_HIDDEN [R0, A0, R1, A1]
      || WIRE [R1, A1]
      end par
   end hide 
end process

------------------------------------------------------------------------
-- shield with two elements (and closed right end)

process SHIELD_2 [R0, A0: LINK] is
   hide R1, A1, R2, A2: LINK in
      par
                 R1, A1 -> SEQUENCER_HIDDEN [R0, A0, R1, A1]
      || R1, A1, R2, A2 -> SEQUENCER_HIDDEN [R1, A1, R2, A2]
      || R2, A2 ->         WIRE [R2, A2]
      end par
   end hide 
end process

------------------------------------------------------------------------
-- shield with five elements (and closed right end)

process SHIELD_5 [R0, A0: LINK] is
   hide R1, A1, R2, A2, R3, A3, R4, A4, R5, A5: LINK in
      par
                 R1, A1 -> SEQUENCER_HIDDEN [R0, A0, R1, A1]
      || R1, A1, R2, A2 -> SEQUENCER_HIDDEN [R1, A1, R2, A2]
      || R2, A2, R3, A3 -> SEQUENCER_HIDDEN [R2, A2, R3, A3]
      || R3, A3, R4, A4 -> SEQUENCER_HIDDEN [R3, A3, R4, A4]
      || R4, A4, R5, A5 -> SEQUENCER_HIDDEN [R4, A4, R5, A5]
      || R5, A5 ->         WIRE [R5, A5]
      end par
   end hide 
end process

------------------------------------------------------------------------
-- shield with ten elements (and closed right end)

process SHIELD_10 [R0, A0: LINK] is
   hide
      R1, A1, R2, A2, R3, A3, R4, A4, R5,  A5,
      R6, A6, R7, A7, R8, A8, R9, A9, R10, A10: LINK
   in
      par
                  R1,  A1  -> SEQUENCER_HIDDEN [R0, A0, R1,  A1]
      || R1,  A1, R2,  A2  -> SEQUENCER_HIDDEN [R1, A1, R2,  A2]
      || R2,  A2, R3,  A3  -> SEQUENCER_HIDDEN [R2, A2, R3,  A3]
      || R3,  A3, R4,  A4  -> SEQUENCER_HIDDEN [R3, A3, R4,  A4]
      || R4,  A4, R5,  A5  -> SEQUENCER_HIDDEN [R4, A4, R5,  A5]
      || R5,  A5, R6,  A6  -> SEQUENCER_HIDDEN [R5, A5, R6,  A6]
      || R6,  A6, R7,  A7  -> SEQUENCER_HIDDEN [R6, A6, R7,  A7]
      || R7,  A7, R8,  A8  -> SEQUENCER_HIDDEN [R7, A7, R8,  A8]
      || R8,  A8, R9,  A9  -> SEQUENCER_HIDDEN [R8, A8, R9,  A9]
      || R9,  A9, R10, A10 -> SEQUENCER_HIDDEN [R9, A9, R10, A10]
      || R10, A10 ->          WIRE [R10, A10]
      end par
   end hide 
end process

end module
\end{lstlisting}

\section{EXP Compositions}
\label{app:exp}
\lstset{language=exp}

This section only presents those composition expressions that cannot be inlined in the SVL script (see Appendix~\ref{app:svl}).

\subsection{Correct Pipelining of Sequencers}
\label{app:exp:pipe}

In the EXP expression corresponding to a correct composition, the names of the two sequencers have to be replaced by concrete file names (see also Sect.~\ref{sec:serial-shield}).
The SVL script (see Appendix~\ref{app:svl}) uses \texttt{sed} to generate appropriate instances.

\begin{lstlisting}
hide R, A in
   par R, A in
      rename R_SUCC -> R, A_SUCC -> A in
         "%%%C1%%%"
      end rename
   || rename R_PRED -> R, A_PRED -> A in
         "%%%C2%%%"
      end rename
   end par
\end{lstlisting}

\noindent
To generate more informative diagnostic traces to the deadlocks, it is better not to hide any gates and to rename the internal gates of the sequencers so as to distinguish between those of the left and right one.

\begin{lstlisting}
par R, A in
   rename
      R_SUCC -> R, A_SUCC -> A, G -> G_L, H -> H_L,
      R_PRED1 -> R_PRED1_L, R_PRED2 -> R_PRED2_L,
      A_SUCC1 -> A_SUCC1_L, A_SUCC2 -> A_SUCC2_L,
      H1 -> H1_L, H2 -> H2_L
   in
      "%%%C1%%%"
   end rename
|| rename
      R_PRED -> R, A_PRED -> A, G -> G_R, H -> H_R,
      R_PRED1 -> R_PRED1_R, R_PRED2 -> R_PRED2_R,
      A_SUCC1 -> A_SUCC1_R, A_SUCC2 -> A_SUCC2_R,
      H1 -> H1_R, H2 -> H2_R
   in
      "%%%C2%%%"
   end rename
end par
\end{lstlisting}

\noindent
Unsurprisingly, this second composition operator yields larger state spaces, e.g.:
\begin{center}
  \small
  \begin{tabular}{lcrrrr}
    model & forks &
    \multicolumn{2}{c}{one sequencer} &
    \multicolumn{2}{c}{two sequencers} \\
    & &
    \multicolumn{1}{c}{states} & \multicolumn{1}{c}{transitions} &
    \multicolumn{1}{c}{states} & \multicolumn{1}{c}{transitions} \\
    \hline
    INTUITIVE  & RV  &   130 &    294 &       651 &       1707 \\
    INTUITIVE  & PPP & 94852 & 357348 & 429321272 & 2443520901 \\
    TRANSITION & PPP & 18136 &  72974 &  28586926 &  189893233 \\
  \end{tabular}
\end{center}

{}

\subsection{Short-Circuits}
\label{app:exp:shortcircuit}

The definition of short-circuits requires synchronization vectors, which are not yet supported inline by SVL.

\subsubsection{Short-Circuit R1-R2}

\begin{lstlisting}
hide R1, A1, R2, A2, R1R2 in
  label par using
     -- synchronization vectors for unmodified wires
     "R_PRED !DOWN" * _          * _              -> "R_PRED !DOWN",
     "R_PRED !UP"   * _          * _              -> "R_PRED !UP",
     "A_PRED !DOWN" * _          * _              -> "A_PRED !DOWN",
     "A_PRED !UP"   * _          * _              -> "A_PRED !UP",
     "A1 !DOWN"     * "A1 !DOWN" * _              -> "A1 !DOWN",
     "A1 !UP"       * "A1 !UP"   * _              -> "A1 !UP",
     _              * "A2 !DOWN" * "A2 !DOWN"     -> "A2 !DOWN",
     _              * "A2 !UP"   * "A2 !UP"       -> "A2 !UP",
     _              * _          * "R_SUCC !DOWN" -> "R_SUCC !DOWN",
     _              * _          * "R_SUCC !UP"   -> "R_SUCC !UP",
     _              * _          * "A_SUCC !DOWN" -> "A_SUCC !DOWN",
     _              * _          * "A_SUCC !UP"   -> "A_SUCC !UP",
     -- synchronization vectors for the short-circuit
     "R1 !DOWN"     * "R1 !DOWN" * "R2 !DOWN"     -> "R1R2 !DOWN",
     "R1 !DOWN"     * "R1 !DOWN" * "R2 !UP"       -> "R1R2 !DOWN",
     "R1 !DOWN"     * "R1 !DOWN" * "R2 !UP"       -> "R1R2 !UP",
     "R1 !UP"       * "R1 !UP"   * "R2 !DOWN"     -> "R1R2 !DOWN",
     "R1 !UP"       * "R1 !UP"   * "R2 !DOWN"     -> "R1R2 !UP",
     "R1 !UP"       * "R1 !UP"   * "R2 !UP"       -> "R1R2 !UP",
     "R1 !DOWN"     * "R2 !DOWN" * "R2 !DOWN"     -> "R1R2 !DOWN",
     "R1 !DOWN"     * "R2 !UP"   * "R2 !UP"       -> "R1R2 !DOWN",
     "R1 !DOWN"     * "R2 !UP"   * "R2 !UP"       -> "R1R2 !UP",
     "R1 !UP"       * "R2 !DOWN" * "R2 !DOWN"     -> "R1R2 !DOWN",
     "R1 !UP"       * "R2 !DOWN" * "R2 !DOWN"     -> "R1R2 !UP",
     "R1 !UP"       * "R2 !UP"   * "R2 !UP"       -> "R1R2 !UP"
  in
     rename R_SUCC -> R1, A_SUCC -> A1 in
        "PROTOCOL.bcg"
     end rename
  || rename R_PRED -> R1, A_PRED -> A1, R_SUCC -> R2, A_SUCC -> A2 in
        "PROTOCOL.bcg"
     end rename
  || rename R_PRED -> R2, A_PRED -> A2 in
        "PROTOCOL.bcg"
     end rename
  end par
\end{lstlisting}

\subsubsection{Short-Circuit R1-A1}

\begin{lstlisting}
hide R1, A1, R2, A2, R1A1 in
  label par using
     -- synchronization vectors for unmodified wires
     "R_PRED !DOWN" * _          * _              -> "R_PRED !DOWN",
     "R_PRED !UP"   * _          * _              -> "R_PRED !UP",
     "A_PRED !DOWN" * _          * _              -> "A_PRED !DOWN",
     "A_PRED !UP"   * _          * _              -> "A_PRED !UP",
     _              * "R2 !DOWN" * "R2 !DOWN"     -> "R2 !DOWN",
     _              * "R2 !UP"   * "R2 !UP"       -> "R2 !UP",
     _              * "A2 !DOWN" * "A2 !DOWN"     -> "A2 !DOWN",
     _              * "A2 !UP"   * "A2 !UP"       -> "A2 !UP",
     _              * _          * "R_SUCC !DOWN" -> "R_SUCC !DOWN",
     _              * _          * "R_SUCC !UP"   -> "R_SUCC !UP",
     _              * _          * "A_SUCC !DOWN" -> "A_SUCC !DOWN",
     _              * _          * "A_SUCC !UP"   -> "A_SUCC !UP",
     -- synchronization vectors for the short-circuit
     "R1 !DOWN"     * "R1 !DOWN" * _              -> "R1A1 !DOWN",
     "R1 !UP"       * "R1 !UP"   * _              -> "R1A1 !UP",
     "R1 !DOWN"     * "A1 !DOWN" * _              -> "R1A1 !DOWN",
     "R1 !DOWN"     * "A1 !UP"   * _              -> "R1A1 !DOWN",
     "R1 !DOWN"     * "A1 !UP"   * _              -> "R1A1 !UP",
     "R1 !UP"       * "A1 !DOWN" * _              -> "R1A1 !DOWN",
     "R1 !UP"       * "A1 !DOWN" * _              -> "R1A1 !UP",
     "R1 !UP"       * "A1 !UP"   * _              -> "R1A1 !UP",
     "A1 !DOWN"     * "R1 !DOWN" * _              -> "R1A1 !DOWN",
     "A1 !DOWN"     * "R1 !UP"   * _              -> "R1A1 !DOWN",
     "A1 !DOWN"     * "R1 !UP"   * _              -> "R1A1 !UP",
     "A1 !UP"       * "R1 !DOWN" * _              -> "R1A1 !DOWN",
     "A1 !UP"       * "R1 !DOWN" * _              -> "R1A1 !UP",
     "A1 !UP"       * "R1 !UP"   * _              -> "R1A1 !UP",
     "A1 !DOWN"     * "A1 !DOWN" * _              -> "R1A1 !DOWN",
     "A1 !UP"       * "A1 !UP"   * _              -> "R1A1 !UP"
  in
     rename R_SUCC -> R1, A_SUCC -> A1 in
        "PROTOCOL.bcg"
     end rename
  || rename R_PRED -> R1, A_PRED -> A1, R_SUCC -> R2, A_SUCC -> A2 in
        "PROTOCOL.bcg"
     end rename
  || rename R_PRED -> R2, A_PRED -> A2 in
        "PROTOCOL.bcg"
     end rename
  end par
\end{lstlisting}

\noindent 
There are no three-party rendezvous: thus, this attack could also be defined for only two sequencers.

\ 
{}

\subsubsection{Short-Circuit R1-A2}

\begin{lstlisting}
hide R1, A1, R2, A2, R1A2 in
  label par using
     -- synchronization vectors for unmodified wires
     "R_PRED !DOWN" * _          * _              -> "R_PRED !DOWN",
     "R_PRED !UP"   * _          * _              -> "R_PRED !UP",
     "A_PRED !DOWN" * _          * _              -> "A_PRED !DOWN",
     "A_PRED !UP"   * _          * _              -> "A_PRED !UP",
     "A1 !DOWN"     * "A1 !DOWN" * _              -> "A1 !DOWN",
     "A1 !UP"       * "A1 !UP"   * _              -> "A1 !UP",
     _              * "R2 !DOWN" * "R2 !DOWN"     -> "R2 !DOWN",
     _              * "R2 !UP"   * "R2 !UP"       -> "R2 !UP",
     _              * _          * "R_SUCC !DOWN" -> "R_SUCC !DOWN",
     _              * _          * "R_SUCC !UP"   -> "R_SUCC !UP",
     _              * _          * "A_SUCC !DOWN" -> "A_SUCC !DOWN",
     _              * _          * "A_SUCC !UP"   -> "A_SUCC !UP",
     -- synchronization vectors for the short-circuit
     "R1 !DOWN"     * "R1 !DOWN" * "A2 !DOWN"     -> "R1A2 !DOWN",
     "R1 !DOWN"     * "R1 !DOWN" * "A2 !UP"       -> "R1A2 !DOWN",
     "R1 !DOWN"     * "R1 !DOWN" * "A2 !UP"       -> "R1A2 !UP",
     "R1 !UP"       * "R1 !UP"   * "A2 !DOWN"     -> "R1A2 !DOWN",
     "R1 !UP"       * "R1 !UP"   * "A2 !DOWN"     -> "R1A2 !UP",
     "R1 !UP"       * "R1 !UP"   * "A2 !UP"       -> "R1A2 !UP",
     "R1 !DOWN"     * "A2 !DOWN" * "A2 !DOWN"     -> "R1A2 !DOWN",
     "R1 !DOWN"     * "A2 !UP"   * "A2 !UP"       -> "R1A2 !DOWN",
     "R1 !DOWN"     * "A2 !UP"   * "A2 !UP"       -> "R1A2 !UP",
     "R1 !UP"       * "A2 !DOWN" * "A2 !DOWN"     -> "R1A2 !DOWN",
     "R1 !UP"       * "A2 !DOWN" * "A2 !DOWN"     -> "R1A2 !UP",
     "R1 !UP"       * "A2 !UP"   * "A2 !UP"       -> "R1A2 !UP"
  in
     rename R_SUCC -> R1, A_SUCC -> A1 in
        "PROTOCOL.bcg"
     end rename
  || rename R_PRED -> R1, A_PRED -> A1, R_SUCC -> R2, A_SUCC -> A2 in
        "PROTOCOL.bcg"
     end rename
  || rename R_PRED -> R2, A_PRED -> A2 in
        "PROTOCOL.bcg"
     end rename
  end par
\end{lstlisting}

\subsubsection{Short-Circuit R2-A1}

\begin{lstlisting}
hide R1, A1, R2, A2, R2A1 in
  label par using
     -- synchronization vectors for unmodified wires
     "R_PRED !DOWN" * _          * _              -> "R_PRED !DOWN",
     "R_PRED !UP"   * _          * _              -> "R_PRED !UP",
     "A_PRED !DOWN" * _          * _              -> "A_PRED !DOWN",
     "A_PRED !UP"   * _          * _              -> "A_PRED !UP",
     "R1 !DOWN"     * "R1 !DOWN" * _              -> "R1 !DOWN",
     "R1 !UP"       * "R1 !UP"   * _              -> "R1 !UP",
     _              * "A2 !DOWN" * "A2 !DOWN"     -> "A2 !DOWN",
     _              * "A2 !UP"   * "A2 !UP"       -> "A2 !UP",
     _              * _          * "R_SUCC !DOWN" -> "R_SUCC !DOWN",
     _              * _          * "R_SUCC !UP"   -> "R_SUCC !UP",
     _              * _          * "A_SUCC !DOWN" -> "A_SUCC !DOWN",
     _              * _          * "A_SUCC !UP"   -> "A_SUCC !UP",
     -- synchronization vectors for the short-circuit
     "A1 !DOWN"     * "A1 !DOWN" * "R2 !DOWN"     -> "A1R2 !DOWN",
     "A1 !DOWN"     * "A1 !DOWN" * "R2 !UP"       -> "A1R2 !DOWN",
     "A1 !DOWN"     * "A1 !DOWN" * "R2 !UP"       -> "A1R2 !UP",
     "A1 !UP"       * "A1 !UP"   * "R2 !DOWN"     -> "A1R2 !DOWN",
     "A1 !UP"       * "A1 !UP"   * "R2 !DOWN"     -> "A1R2 !UP",
     "A1 !UP"       * "A1 !UP"   * "R2 !UP"       -> "A1R2 !UP",
     "A1 !DOWN"     * "R2 !DOWN" * "R2 !DOWN"     -> "A1R2 !DOWN",
     "A1 !DOWN"     * "R2 !UP"   * "R2 !UP"       -> "A1R2 !DOWN",
     "A1 !DOWN"     * "R2 !UP"   * "R2 !UP"       -> "A1R2 !UP",
     "A1 !UP"       * "R2 !DOWN" * "R2 !DOWN"     -> "A1R2 !DOWN",
     "A1 !UP"       * "R2 !DOWN" * "R2 !DOWN"     -> "A1R2 !UP",
     "A1 !UP"       * "R2 !UP"   * "R2 !UP"       -> "A1R2 !UP"
  in
     rename R_SUCC -> R1, A_SUCC -> A1 in
        "PROTOCOL.bcg"
     end rename
  || rename R_PRED -> R1, A_PRED -> A1, R_SUCC -> R2, A_SUCC -> A2 in
        "PROTOCOL.bcg"
     end rename
  || rename R_PRED -> R2, A_PRED -> A2 in
        "PROTOCOL.bcg"
     end rename
  end par
\end{lstlisting}

\subsubsection{Short-Circuit R2-A2}

As for the short-circuit \lstinline+R1+-\lstinline+A1+, there are no three-party rendezvous.

\begin{lstlisting}
hide R1, A1, R2, A2, R2A2 in
  label par using
     -- synchronization vectors for unmodified wires
     "R_PRED !DOWN" * _          * _              -> "R_PRED !DOWN",
     "R_PRED !UP"   * _          * _              -> "R_PRED !UP",
     "A_PRED !DOWN" * _          * _              -> "A_PRED !DOWN",
     "A_PRED !UP"   * _          * _              -> "A_PRED !UP",
     "R1 !DOWN"     * "R1 !DOWN" * _              -> "R1 !DOWN",
     "R1 !UP"       * "R1 !UP"   * _              -> "R1 !UP",
     "A1 !DOWN"     * "A1 !DOWN" * _              -> "A1 !DOWN",
     "A1 !UP"       * "A1 !UP"   * _              -> "A1 !UP",
     _              * _          * "R_SUCC !DOWN" -> "R_SUCC !DOWN",
     _              * _          * "R_SUCC !UP"   -> "R_SUCC !UP",
     _              * _          * "A_SUCC !DOWN" -> "A_SUCC !DOWN",
     _              * _          * "A_SUCC !UP"   -> "A_SUCC !UP",
     -- synchronization vectors for the short-circuit
     _              * "R2 !DOWN" * "R2 !DOWN"     -> "R2A2 !DOWN",
     _              * "R2 !UP"   * "R2 !UP"       -> "R2A2 !UP",
     _              * "R2 !DOWN" * "A2 !DOWN"     -> "R2A2 !DOWN",
     _              * "R2 !DOWN" * "A2 !UP"       -> "R2A2 !DOWN",
     _              * "R2 !DOWN" * "A2 !UP"       -> "R2A2 !UP",
     _              * "R2 !UP"   * "A2 !DOWN"     -> "R2A2 !DOWN",
     _              * "R2 !UP"   * "A2 !DOWN"     -> "R2A2 !UP",
     _              * "R2 !UP"   * "A2 !UP"       -> "R2A2 !UP",
     _              * "A2 !DOWN" * "R2 !DOWN"     -> "R2A2 !DOWN",
     _              * "A2 !DOWN" * "R2 !UP"       -> "R2A2 !DOWN",
     _              * "A2 !DOWN" * "R2 !UP"       -> "R2A2 !UP",
     _              * "A2 !UP"   * "R2 !DOWN"     -> "R2A2 !DOWN",
     _              * "A2 !UP"   * "R2 !DOWN"     -> "R2A2 !UP",
     _              * "A2 !UP"   * "R2 !UP"       -> "R2A2 !UP",
     _              * "A2 !DOWN" * "A2 !DOWN"     -> "R2A2 !DOWN",
     _              * "A2 !UP"   * "A2 !UP"       -> "R2A2 !UP"
  in
     rename R_SUCC -> R1, A_SUCC -> A1 in
        "PROTOCOL.bcg"
     end rename
  || rename R_PRED -> R1, A_PRED -> A1, R_SUCC -> R2, A_SUCC -> A2 in
        "PROTOCOL.bcg"
     end rename
  || rename R_PRED -> R2, A_PRED -> A2 in
        "PROTOCOL.bcg"
     end rename
  end par
\end{lstlisting}

\subsubsection{Short-Circuit A1-A2}

\begin{lstlisting}
hide R1, A1, R2, A2, A1A2 in
  label par using
     -- synchronization vectors for unmodified wires
     "R_PRED !DOWN" * _          * _              -> "R_PRED !DOWN",
     "R_PRED !UP"   * _          * _              -> "R_PRED !UP",
     "A_PRED !DOWN" * _          * _              -> "A_PRED !DOWN",
     "A_PRED !UP"   * _          * _              -> "A_PRED !UP",
     "R1 !DOWN"     * "R1 !DOWN" * _              -> "R1 !DOWN",
     "R1 !UP"       * "R1 !UP"   * _              -> "R1 !UP",
     _              * "R2 !DOWN" * "R2 !DOWN"     -> "R2 !DOWN",
     _              * "R2 !UP"   * "R2 !UP"       -> "R2 !UP",
     _              * _          * "R_SUCC !DOWN" -> "R_SUCC !DOWN",
     _              * _          * "R_SUCC !UP"   -> "R_SUCC !UP",
     _              * _          * "A_SUCC !DOWN" -> "A_SUCC !DOWN",
     _              * _          * "A_SUCC !UP"   -> "A_SUCC !UP",
     -- synchronization vectors for the short-circuit
     "A1 !DOWN"     * "A1 !DOWN" * "A2 !DOWN"     -> "A1A2 !DOWN",
     "A1 !DOWN"     * "A1 !DOWN" * "A2 !UP"       -> "A1A2 !DOWN",
     "A1 !DOWN"     * "A1 !DOWN" * "A2 !UP"       -> "A1A2 !UP",
     "A1 !UP"       * "A1 !UP"   * "A2 !DOWN"     -> "A1A2 !DOWN",
     "A1 !UP"       * "A1 !UP"   * "A2 !DOWN"     -> "A1A2 !UP",
     "A1 !UP"       * "A1 !UP"   * "A2 !UP"       -> "A1A2 !UP",
     "A1 !DOWN"     * "A2 !DOWN" * "A2 !DOWN"     -> "A1A2 !DOWN",
     "A1 !DOWN"     * "A2 !UP"   * "A2 !UP"       -> "A1A2 !DOWN",
     "A1 !DOWN"     * "A2 !UP"   * "A2 !UP"       -> "A1A2 !UP",
     "A1 !UP"       * "A2 !DOWN" * "A2 !DOWN"     -> "A1A2 !DOWN",
     "A1 !UP"       * "A2 !DOWN" * "A2 !DOWN"     -> "A1A2 !UP",
     "A1 !UP"       * "A2 !UP"   * "A2 !UP"       -> "A1A2 !UP"
  in
     rename R_SUCC -> R1, A_SUCC -> A1 in
        "PROTOCOL.bcg"
     end rename
  || rename R_PRED -> R1, A_PRED -> A1, R_SUCC -> R2, A_SUCC -> A2 in
        "PROTOCOL.bcg"
     end rename
  || rename R_PRED -> R2, A_PRED -> A2 in
        "PROTOCOL.bcg"
     end rename
  end par
\end{lstlisting}

\section{SVL Script}
\label{app:svl}
\lstset{language=svl}

The following SVL script generates the expected behavior, performs the circuit-level analysis of the various attacks, and attempts the generation of a pipe-line of two sequencers modeled at gate level, for each of the various models of wires, forks, and gates.

Running this SVL script on a laptop with an Intel\textsuperscript{\textregistered} Core\texttrademark i5 M560 CPU at 2.67~GHz, 8~GB of RAM, Debian GNU/Linux~9, and CADP~2019-k (Pisa) in 32-bit mode takes about 15~minutes.

\begin{lstlisting}[firstline=5]
% DEFAULT_PROCESS_FILE="sequencer.lnt"

------------------------------------------------------------------------
-- choice of an implementation of gates

% SET_MODEL () {
%	# "$1" should be one of:
%	# FREE, INTUITIVE, PARALLEL, STATE, TRANSITION
%	rm -f GATES.lnt
%	$CADP/src/com/cadp_ln ${1}_GATES.lnt GATES.lnt
% }
% SVL_RECORD_FOR_CLEAN GATES.lnt
% SET_MODEL TRANSITION

------------------------------------------------------------------------
-- pipelined parallel composition of two sequencers

% PIPE () {
% # $1 : C1 (composant a gauche)
% # $2 : C2 (composant a droite)
% # $3 : C1 ||_{R,A} C2

	% sed -e "s/%%%C1%%%/$1.bcg/" -e "s/%%%C2%%%/$2.bcg/" \
	% composition.exp > ${SVL_TMP_PREFIX}_composition.exp
	% SVL_RECORD_FOR_SWEEP ${SVL_TMP_PREFIX}_composition.exp

	"$3.bcg" =
		divbranching reduction of
		"${SVL_TMP_PREFIX}_composition.exp" ;
% }

------------------------------------------------------------------------
-- circuit-level analysis
------------------------------------------------------------------------

-- generation of shields with one, two, and three sequencers

"PROTOCOL.bcg" = reduction of "PROTOCOL" ;
% PIPE PROTOCOL PROTOCOL PROTOCOL_PROTOCOL
% PIPE PROTOCOL PROTOCOL_PROTOCOL PROTOCOL_PROTOCOL_PROTOCOL

------------------------------------------------------------------------

property INDUCTION_BASIS
	"PROTOCOL || PROTOCOL = PROTOCOL"
	"PROTOCOL || PROTOCOL || PROTOCOL = PROTOCOL"
is
	comparison "PROTOCOL_PROTOCOL.bcg" == "PROTOCOL.bcg" ;
	expected TRUE ;
	comparison "PROTOCOL_PROTOCOL_PROTOCOL.bcg" == "PROTOCOL.bcg" ;
	expected TRUE ;
end property

------------------------------------------------------------------------

"STUCKAT_DOWN.bcg" = reduction of "STUCKAT [W] (DOWN)" ;
"STUCKAT_UP.bcg" = reduction of "STUCKAT [W] (UP)" ;

------------------------------------------------------------------------

property STUCKAT (WIRE, VOLTAGE)
	"wire $WIRE stuck at $VOLTAGE"
is
	-- both sequencers have to synchronize on $WIRE at $VOLTAGE
	branching comparison
		hide R, A in
			par
			   R, A ->
				rename R_SUCC -> R, A_SUCC -> A in
					"PROTOCOL.bcg"
				end rename
			|| R, A ->
				rename R_PRED -> R, A_PRED -> A in
					"PROTOCOL.bcg"
				end rename
			|| "$WIRE" ->
				rename W -> "$WIRE" in
					"STUCKAT_$VOLTAGE.bcg"
				end rename
			end par
		end hide
	>=
	"PROTOCOL.bcg" ;
	expected FALSE ;

	-- the emitting sequencer can put any voltage on $WIRE
	-- the receiving sequencer can only get $VOLTAGE on $WIRE
	% if [ $WIRE = R ] ; then
	%	OTHER_WIRE=A
	%	SEND=SUCC
	%	RECV=PRED
	% else
	%	OTHER_WIRE=R
	%	SEND=PRED
	%	RECV=SUCC
	% fi
	branching comparison
		hide R, A in
			par
			   "$OTHER_WIRE" ->
				rename "R_$SEND" -> R, "A_$SEND" -> A in
					"PROTOCOL.bcg"
				end rename
			|| R, A ->
				rename "R_$RECV" -> R, "A_$RECV" -> A in
					"PROTOCOL.bcg"
				end rename
			|| "$WIRE" ->
				rename W -> "$WIRE" in
					"STUCKAT_$VOLTAGE.bcg"
				end rename
			end par
		end hide
	>=
	"PROTOCOL.bcg" ;
	expected FALSE ;
end property

check STUCKAT (R, DOWN) ;
check STUCKAT (R, UP) ;
check STUCKAT (A, DOWN) ;
check STUCKAT (A, UP) ;

------------------------------------------------------------------------

property CUT (WIRE)
	"wire $WIRE cut"
	"(generates warnings about a missing synchronization)"
is
	-- both sequencers cannot synchronize on $WIRE
	branching comparison
		hide R, A in
			par
			   R, A ->
				rename R_SUCC -> R, A_SUCC -> A in
					"PROTOCOL.bcg"
				end rename
			|| R, A ->
				rename R_PRED -> R, A_PRED -> A in
					"PROTOCOL.bcg"
				end rename
			|| "$WIRE" ->
				rename W -> "$WIRE" in
					stop
				end rename
			end par
		end hide
	>=
	"PROTOCOL.bcg" ;
	expected FALSE ;

	-- the emitting sequencer can synchronize on $WIRE
	-- the receiving sequencer cannot synchronize on $WIRE
	% if [ $WIRE = R ] ; then
	%	OTHER_WIRE=A
	%	SEND=SUCC
	%	RECV=PRED
	% else
	%	OTHER_WIRE=R
	%	SEND=PRED
	%	RECV=SUCC
	% fi
	branching comparison
		hide R, A in
			par
			   "$OTHER_WIRE" ->
				rename "R_$SEND" -> R, "A_$SEND" -> A in
					"PROTOCOL.bcg"
				end rename
			|| R, A ->
				rename "R_$RECV" -> R, "A_$RECV" -> A in
					"PROTOCOL.bcg"
				end rename
			|| "$WIRE" ->
				rename W -> "$WIRE" in
					stop
				end rename
			end par
		end hide
	>=
	"PROTOCOL.bcg" ;
	expected FALSE ;

	-- both sequencer can synchronize on $WIRE,
	-- but without synchronizing with each other
	-- this kind of wire cut is unrealistic and not detected
	branching comparison
		hide R, A in
			par "$OTHER_WIRE" in
				rename R_SUCC -> R, A_SUCC -> A in
					"PROTOCOL.bcg"
				end rename
			||	rename R_PRED -> R, A_PRED -> A in
					"PROTOCOL.bcg"
				end rename
			end par
		end hide
	>=
	"PROTOCOL.bcg" ;
	expected TRUE ;
end property

check CUT (R) ;
check CUT (A) ;

------------------------------------------------------------------------

property SHORTCIRCUIT (WIRE1, WIRE2, RESULT)
	"shortcircuit between $WIRE1 and $WIRE2"
is
	"composition_shortcircuit_$WIRE1$WIRE2.bcg" =
		branching reduction of
		"composition_shortcircuit_$WIRE1$WIRE2.exp" ;
	branching comparison
		"composition_shortcircuit_$WIRE1$WIRE2.bcg"
	>=
	"PROTOCOL.bcg" ;
	expected "$RESULT" ;
end property

check SHORTCIRCUIT (R1, R2, FALSE) ;
check SHORTCIRCUIT (R1, A1, TRUE) ;
check SHORTCIRCUIT (R1, A2, FALSE) ;
check SHORTCIRCUIT (R2, A1, FALSE) ;
check SHORTCIRCUIT (R2, A2, TRUE) ;
check SHORTCIRCUIT (A1, A2, FALSE) ;

------------------------------------------------------------------------
-- gate-level analysis
------------------------------------------------------------------------

"STUB_L.bcg" = divbranching reduction of "STUB_L" ;
"STUB_R.bcg" = divbranching reduction of "STUB_R" ;

% for MODEL in TRANSITION INTUITIVE FREE STATE PARALLEL ; do
%	SET_MODEL $MODEL
%	for FORKS in RV PPP PPI PIP PII IPP IPI IIP III ; do

		"$MODEL.$FORKS.bcg" =
			divbranching reduction of
			gate hide all but R_PRED, A_PRED, R_SUCC, A_SUCC
			in "SEQUENCER_${FORKS} (DOWN, DOWN, DOWN)" ;

%		SEQ=$MODEL.$FORKS

%		PIPE STUB_L $SEQ L_$SEQ
%		PIPE L_$SEQ STUB_R L_${SEQ}_R

		branching comparison
			"L_${SEQ}_R.bcg" == "PROTOCOL.bcg" ;

		deadlock of branching reduction of "L_${SEQ}_R.bcg" ;

%		if [ $FORKS != RV ] ; then
			-- avoid the generation of large models
%			continue
%		fi
%		PIPE $SEQ $SEQ ${SEQ}_${SEQ}

		deadlock of branching reduction of "${SEQ}_${SEQ}.bcg" ;
%	done

% done

-------------------------------------------------------------------------
\end{lstlisting}

\end{document}